
\magnification=1200
\font\gran=cmbx12
\line{}
\line{\hfil UB-ECM-PF-94/24}
\line{\hfil August 1994}
\bigskip\bigskip\bigskip\bigskip
\centerline{{\gran Perturbative QCD}
	 \footnote{$^\dagger$}{
Lectures delivered at the XXII International Meeting on Fundamental
 Physics ``The Standard Model and Beyond", Jaca, February 1994. To be
published in the proceedings.}}

\bigskip\bigskip
\centerline{ D. Espriu\footnote{$^*$}{E-mail: espriu@ebubecm1.bitnet,
espriu@greta.ecm.ub.es}  }
\bigskip
\centerline{\it D.E.C.M., Facultat de F\'\i sica and I.F.A.E.}
\centerline{\it Universitat de Barcelona}
\centerline{\it Diagonal 647}
\centerline{\it E-08028 Barcelona}
\bigskip\bigskip\bigskip\bigskip
\centerline{ABSTRACT}
\medskip
This is the written version of a set of lectures on perturbative
QCD that were delivered to a mixed audience of young
theorists and experimentalists in the course of the XXII International
Meeting on Fundamental Physics.
These notes are virtually a verbatim transcription of the
lectures.  The selection of  topics
is somewhat arbitrary, but
two basic points are emphasized:
the rationale behind QCD
and how ongoing experiments, such as those
taking place in LEP and HERA,  contribute to our
understanding of strong interactions. The lectures are thus
roughly divided into three parts: foundations of QCD,
determination of $\alpha_s$ and jet physics, and
deep inelastic experiments.

\vfill
\eject

\line{\bf 1.- Introduction\hfil}
\medskip\noindent
It is not the purpose of these notes to give a detailed
account of Quantum Chromodynamics, the theory of strong
interactions. There are many good textbooks$^{1)}$ that
are excellent at doing this job.
We do not pretend  to treat exhaustively
any particular aspect of Quantum Chromodynamics  or attempt to
provide a general review of
the status of this theory, either.

These lectures are essentially addressed to young experimentalists
without too much theoretical background in
Quantum  Field Theory.
We have attempted to single
out amongst the theoretical foundations of QCD
the ones where Quantum Chromodynamics really hinges on, trying to
present them in terms as simple and physical as possible.
 Then we
 move to the connection between theory and experiment. This
is not a simple problem in QCD. It is well known that the
fields and particles  we know how to compute with
(with the simplest tool at our disposal, perturbation
theory) are not those that are observed by experimentalists
in their detectors due to the phenomenon of confinement. Quarks and gluons
are real, but they cannot be detected as free particles as they are believed
to be confined inside hadrons. In view of this is quite remarkable
that there are  theoretical techniques enabling us to put the
theory
to very stringent tests. To mention just a significative measurement, the
value $\alpha_s(M_Z^2)=0.123\pm
0.006$ from an analysis of the process $e^+e^- \to \bar{q} q$ has been
reported in this meeting$^{2)}$. This is an impressive
accuracy for a number that is not, strictly speaking, observable
due to confinement. Even more so when we compare this precision
to the one
we had only a few years
ago. Reaching
this level of accuracy is one of the achievements of LEP.

Perturbative QCD can be applied to inclusive processes provided
that the characteristic momentum transfer is large enough.
In these lectures we will discuss is some detail
these techniques and apply them to the determination of $\alpha_s$
through $R_{had}$
$R_\tau$, or from jet topology.
Another test of perturbative
QCD is, of course, deep inelastic scattering. Here the commissioning
of HERA has opened a new kinematical region where it will be
possible to study the onset of non-perturbative effects.
The exploration of this region is a fascinating subject
interesting on its own right.

Due to the lack of space we have not included two sections that, in our
view, should be in any general review of QCD. The first one
concerns the
so-called ``spin of the proton" problem$^{3)}$. Another topic that is not
covered at all is the study of the photon structure
functions. We direct the interested reader to ref. $^{4)}$.

We have not really made any serious attempt to provide
a complete set of references. Those provided  merely reflect
personal tastes. A more complete bibliography can be found
in some of the references, as indicated.
\bigskip

\line{\bf 2.- Why QCD\hfil}
\medskip\noindent
Since Dirac we know that the vacuum in Quantum
Field Theory is a rather peculiar object. The vacuum is
defined by the property that all its negative energy
states are filled, while those with positive energy
are empty. A virtual photon with sufficient energy
will promote a negative state to a positive one, thus
creating a particle-hole (electron-positron)
pair. Eventually, the positive energy state will
decay filling again the hole in the Dirac sea
with the emission of a off-shell photon which, in turn, will excite another
pair and so on. This is illustrated in fig. 1.

\vbox{\bigskip
\vskip 3.5truecm
\medskip
\centerline{Fig. 1.- The Dirac sea.}
\bigskip}

In the quantum world this process takes place
virtually all the
time.  The uncertainty relation
$\Delta E \Delta t > \hbar$
allows that for a short time
a pair is created, then annihilated. The vacuum is
thus constantly populated by virtual electron-positron
pairs, becoming a dielectric medium at quantum scales.
A medium that partially screens charges. The screening
is more and more effective at larger distances since more and more
virtual pairs enter the game. Therefore
$$ \alpha\to \alpha_{eff}(r) $$
$$\alpha_{eff}(r)= {{\alpha}\over{1+{2\over{ 3\pi}}\alpha
\log{r\over r_0}}},
\eqno(1)$$
$r_0$ is some scale at which we choose to measure $\alpha$.
At large scales the effective electromagnetic coupling
decreases, becoming larger at short distances. While it
requires some work to get the detailed form of eq. 1, the
reasoning behind the behaviour of the effective charge
for $r\gg r_0$ is so general that it is difficult
to see how it could go otherwise.

Yet, the opposite behaviour takes place in Quantum
Chromodynamics. Fig. 2 shows the elastic and total
cross sections for $\pi$ scattering off protons.
The cross sections show a rather complicate resonant behaviour at
low energies but become much simpler at high energies.
This suggests that strong interactions become
stronger at low energies, i.e. long distances.

\vbox{\bigskip
\vskip 4.5truecm
\medskip
\centerline{Fig. 2.- A typical hadronic cross-section at low
energies$^{5)}$} \bigskip}

In a completely different regime
deep inelastic scattering (fig. 3)
showed  25
years ago the existence of hard processes inside
nucleons, at very short distances. One can get a rough picture
of these processes by just assuming that quarks are
approximately free if one looks at  distances
$\ll 1$ (GeV)$^{-1}$.

\vbox{\bigskip
\vskip 3truecm
\medskip
\centerline{Fig. 3.- The kinematics of Deep Inelastic Scattering.}
\bigskip}

One needs  a rather peculiar type of theory. It
must be a theory whose elementary fields are fermions
(quarks), but weakly interacting
at short distances. In addition we would like it to
be a renormalizable theory to be able to apply the full machinery
 Quantum Field Theory
and this pretty much
forces us to work with gauge theories$^{6)}$. Trivial modifications
of QED will not work because we need  anti-screenning of
charges rather than screening.

Several hints as to the way to go come from the successful
quark model. We have for instance the $\Delta^{++}
= \vert u^\uparrow u^\uparrow u^\uparrow\rangle $. Being
the lightest hadron with this quark contents we expect to have
the three quarks in the ground state, hence in a symmetric wave function.
This is in contradiction with Fermi statistics. The contradiction can be
solved if we admit the existence of a new quantum number $\alpha$ and
$$\vert u^\uparrow u^\uparrow u^\uparrow\rangle
= {1\over \sqrt{6}}\epsilon^{\alpha\beta\gamma}
\vert u_\alpha^\uparrow u_\beta^\uparrow u_\gamma^\uparrow\rangle.
\eqno(2)$$
Indications in the same direction also come from two very different
processes. One is the cross-section for $e^+ e^-\to hadrons$.
Assuming a free quark model (which as we mentioned
should be good as the momentum transfer is large) we have
$$ R_{had}={{\sigma(e^+ e^-\to hadrons)}\over
{\sigma(e^+ e^- \to \mu^+\mu^-)}}
\sim \sum_{i=1}^n Q_i^2. \eqno(3)$$
The index $i$ runs over all degrees of freedom
coupling to the intermediate photon (with charge $Q_i$).
$$n=3 \qquad u,d,s\phantom{,c,b,t}\qquad \,R={2\over 3} $$
$$n=4 \qquad u,d,s,c\phantom{,b,t}\qquad R={10\over 9}\eqno(4)$$
$$n=6 \qquad u,d,s,c,b,t\qquad \,R={5\over 3}. $$
Experimentally all these values are off by about a factor 3.
The quark model can again be reconciled with experiment if we
admit that the new quantum number can take 3 values
and is blind to electromagnetic interactions,
$q^\alpha$, $\alpha=1,2,3$.

\vbox{\bigskip
\vskip 2.5 truecm
\medskip
\centerline{Fig. 4.- The decay $\pi^0\to \gamma\gamma$.}
\bigskip}

The other process relevant at this point is the
celebrated $\pi^0\to \gamma\gamma$ decay.
This process goes dominantly through the
triangle diagram shown in fig.4, with a closed
loop of quarks. This diagram is also called
`the anomaly', for reasons that we will discuss
in a while. The calculation gives$^{7)}$
$$\Gamma={1\over{576\pi^3}}
{\alpha^2\over f_\pi^2} m_\pi^3 = 0.85 {\rm eV}.
\eqno(5)$$
Experimentally $\Gamma=7.37\pm 0.5$ eV. The result
is off by a factor $ 9=3^2$ which, again, is understood
if we accept that additional degrees of freedom, invisible to both
 photon and  pion, exist. So we learn that ordinary hadrons seem
not to carry any color. Similar analysis show that color does not
couple to the $W$ or $Z$ bosons either.

The guess
 is now more or less obvious. Let's introduce
a gauge symmetry acting on the new degree of freedom
just as Electromagnetism acts on the electric charge
with a gauge group $U(1)$. It must be a new gauge group, since
color does not couple to known gauge fields.
We want a group with irreducible representations of
dimension 3, for the quark model to fit
in. Obvious candidates are
$O(3)$, $SU(2)$, $U(2)$, $U(3)$, $SU(3)$,
$SO(3)$ and $Sp(2)$. Groups such as
$U(2)$, $U(3)$ and $O(3)$ can be discarded right away
because $\epsilon^{\alpha\beta\gamma}$ is not
an invariant tensor (and we need that for the
$\Delta^{++}$). On the other hand
$Sp(2)\simeq SO(3)= SU(2)/Z_2$. Neither of these
have {\it complex} representations of dimension 3
(hence they would lead to diquark states). We are
left with $SU(3)$.
\bigskip
\line{\bf 3.- Lagrangian and Symmetries\hfil}
\medskip\noindent
The QCD lagrangian is
$${\cal L}_{QCD}=
-{1\over 4}F_{\mu\nu}^a F^{\mu\nu a}
+i\sum_{j=1}^n \bar{\psi}^\alpha_j \gamma^\mu D_{\mu\,\alpha\beta}
\psi^\beta_j -\sum_{j=1}^n m_j\bar{\psi}^\alpha_j \psi_{j\alpha},
\eqno(6)
$$
where
$$F_{\mu\nu}^a=\partial_\mu W_\nu^a -\partial_\nu W_\mu^a
+g f_{abc} W_\mu^b W_\nu^c, \eqno(7)$$
$$ D_{\mu\,\alpha\beta}=\delta_{\alpha\beta}\partial_\mu
-ig{T^a_{\alpha\beta}} W_\mu^a.\eqno(8)$$
The generators $T^a$ are related to the well-known Gell-Mann matrices
and act
on the fundamental
representation of $SU(3)$
$$T^a_{\alpha\beta}={\lambda^a_{\alpha\beta}\over 2}
\qquad [T^a,T^b]=if_{abc}T^c\eqno(9)$$
With the structure constants $f_{abc}$ one constructs the
generators of the adjoint representation
$ T^a_{bc}= if_{abc}$.

${\cal L}_{QCD}$ has a local gauge
invariance. If $G(x)$ is a $SU(3)$ matrix, the transformation
$$ \psi(x)\to G(x)\psi(x)
\qquad W_\mu(x)\to G(x)W_\mu(x)G(x)^{-1}
+{i\over g} \partial_\mu G(x) G(x)^{-1}\eqno(10)$$
leaves ${\cal L}_{QCD}$ invariant. This symmetry
is crucial to remove two of the four
degrees of freedom in the field $W_\mu$,
$(\mu=0,1,2,3)$.

To quantize the theory one must select
a gauge. The bilinear part in the $W_\mu$
field is
$${1\over 2} W_\mu^a(k^2 g^{\mu\nu}-k^\mu k^\nu) W_\nu^a
\equiv {1\over 2} W_\mu^a M^{\mu\nu} W_\nu^a;\eqno(11)$$
$M^{\mu\nu}$ cannot be inverted to find the
propagator. The way out is to add
the piece
$$ {-1\over {2\xi}} (\partial^\mu W_\mu^a)^2.
\eqno(12)$$
Then
$$M^{\mu\nu}=k^2 g^{\mu\nu} -(1-{1\over \xi})k^\mu k^\nu
\qquad
(M^{-1})^{\mu\nu}={{g^{\mu\nu}-(1-\xi){{k^\mu k^\nu}\over k^2}}
\over {k^2 + i\epsilon}}.\eqno(13)$$
We can now write Feynman diagrams. The added term
eq. 12  breaks the local gauge
symmetry which, generally speaking, is only recovered
for $S$-matrix elements.

\vbox{\bigskip
\vskip 2.5truecm
\medskip
\centerline{Fig. 5.- Interaction vertices in QCD. Only the
color factors are shown.}
\bigskip}

${\cal L}_{QCD}$ has several types of interaction
vertices. They are shown in fig. 5. Let us now
consider the process
 $\bar{q} q\to gg$. At tree level the appropriate
diagrams are shown in fig. 6.

\vbox{\bigskip
\vskip 2.5truecm
\medskip
\centerline{Fig. 6.- Tree level diagrams for $\bar{q} q\to gg$.}
\bigskip}

\noindent
(c) is absent in the analogous QED process $e^+e^-\to \gamma\gamma$. Due to
 this diagram it turns out that the above process
has a bad high-energy behaviour when we sum over final state polarizations
covariantly
$$ P={1\over 2}g_{\mu_1\nu_1} g_{\mu_2\nu_2}
J^{\mu_1\mu_2} (J^{\nu_1\nu_2})^\dagger,\eqno(14)$$
but it is just fine if we include transverse
polarizations only
$$P={1\over 2}\tau_{\mu_1\nu_1} \tau_{\mu_2\nu_2}
J^{\mu_1\mu_2} (J^{\nu_1\nu_2})^\dagger.\eqno(15)$$
The tensors
$g_{\mu\nu}$ and $\tau_{\mu\nu}$ are obtained by summing
the polarization vectors $\epsilon_\mu(\sigma)\epsilon_\nu(\sigma)$
over all polarizations or over physical ones only, respectively.
If we insist in keeping a covariant formalism in which we sum over
all four polarizations something must cancel this undesirable
high-energy behaviour. This is accomplished
by adding to ${\cal L}_{QCD}$ the piece
$$ -\partial^\mu \bar{\varphi}_a D_\mu^{ab} \varphi_b\eqno(16)$$
with $D_\mu^{ab}$ being the same as $D_\mu^{\alpha\beta}$
in eq. 8, but with the generators of the fundamental representation
of $SU(3)$ replaced by those of the adjoint one. The fields
$\varphi^a$ have boson-like couplings, but are defined to have Fermi
statistics. They contribute with a (-1) factor to the cross-section.
They are not required in abelian theories like QED, but are crucial in
QCD.

\vbox{\bigskip
\vskip 2 truecm
\medskip
\centerline{Fig. 7.- Ghost contribution to $\bar{q} q\to gg$ cross-section.}
\bigskip}

\noindent
If all polarizations, physical and unphysical, are summed
over one must accept that $\varphi$ states can be produced
even if they are ghost states with unphysical statistics.
The contribution from the piece  we have added to the
lagrangian, eq. 16,
is just right to reproduce the
same results we would get keeping the physical
polarizations only. In practice, in internal loops we have
really no choice but to keep the covariant sum over
polarizations and ghosts have to be included there.
It is also possible to derive the need for the
introduction of ghosts from more formal arguments$^{8)}$, but
that would take us too far afield.

In addition to the {\it local} gauge symmetry we also have
 exact or approximate {\it global} symmetries in
${\cal L}_{QCD}$. The lagrangian is invariant
under the global transformation
$$\psi(x)\to \exp(-i\theta I)\psi(x),\eqno(17)$$
leading to baryon number conservation
$$B=\int d^3x J_0.\eqno(18)$$
If all quark masses are equal there
are additional symmetries and conserved currents$^{9)}$
$$\psi(x)\to \exp(-i\theta^a T^a)\psi(x);\eqno(19)$$
$$J_\mu=\bar{\psi}\gamma_\mu T^a \psi. \eqno(20)$$
$\psi$ now represents a column vector
containing all flavours and $T^a$ is an $SU(N_f)$ generator.
Furthermore, if all quark masses vanish ${\cal L}_{QCD}$
is also invariant under
$$\psi(x)\to \exp(-i\theta^a T^a \gamma_5)\psi(x).\eqno(21)$$
$$J_\mu=\bar{\psi}\gamma_\mu\gamma_5 T^a \psi\eqno(22)$$
are the corresponding conserved currents.
In the real world quark masses are not equal, let alone zero.
The two latter symmetries eqs. 19, 21 are only approximate
and this only for light quarks. Therefore the
hadronic world is {\it approximately} invariant under
$U(1)\times SU(3)_V\times SU(3)_A$. $U(1)$ is always
exact and $SU(3)_V$ is nothing but the
vintage $SU(3)$ of Gell-Mann$^{10)}$, which led to the
quark model thirty years ago.

${\cal L}_{QCD}$ is also invariant under
$$\psi(x)\to \exp(-i\theta I \gamma_5)\psi(x)\eqno(23)$$
The `conserved' current is
$$J^5_\mu=\bar{\psi}\gamma_\mu \gamma^5\psi.\eqno(24)$$
However, when one computes Green functions with insertions
of the divergence of the above current, $\partial^\mu J_\mu^5$,
one gets non-zero answers. The culprit is the
triangle diagram, the `anomaly' (the same that we met
in $\pi^0$ decay, but with the two external photons replaced
by gluons). In fact, a careful calculation shows that
while the axial current (24) is conserved at tree level,
quantum corrections spoil that conservation and, in fact,
$$ \partial_\mu (\bar{\psi}\gamma^\mu\gamma^5\psi)
={g^2\over {4\pi^2}}{N_f\over 8}\epsilon^{\mu\nu\alpha\beta}
F_{\mu\nu}^a F_{\alpha\beta}^a.\eqno(25)$$
The key point is that there is no way of estimating the divergent momentum
integral that appears in the evaluation of fig. 4 without breaking
the $U(1)_A$ symmetry, and this is not a point  of mathematical finesse;
it has far reaching consequences.
The r.h.s. of eq. 25 is itself a total divergence
$\partial^\mu K_\mu$. Then the charge
$Q_5$ verifies
$$\dot{Q}_5=\int d^3x \partial_0 J^0_5 =
\int d^3x \partial_\mu K^\mu
-\int d^3x \partial_i J^i_5\eqno(26)$$
In perturbation theory all fields are
small perturbations from the vacuum;
they decay fast enough to infinity to be
able to neglect all boundary terms in the integrals.
By Gauss theorem the second integral on the r.h.s.
of eq. 26 drops and
$$Q_5(t=+\infty)-Q_5(t=-\infty)=
\int d^4x \partial_\mu K^\mu=0\eqno(27)$$
In a non-abelian gauge theory such as QCD there
are, however, some non-perturbative gauge
configurations$^{11)}$ (i.e. configurations which are
not a superposition of states with a finite number
of quarks and gluons) that do not possess
the nice long distance behaviour that is required
for eq. 27 to vanish. These make $\dot{Q}_5\neq 0$.
A conserved charge can still be defined by the
integral of $J^5_0-K_0$, but it is not gauge
invariant (the divergence of $K_\mu$ is
invariant, but not $K_\mu$ itself). There is no way
to have a conserved, gauge invariant axial charge
in QCD. $U(1)_A$ is not a symmetry of the theory.

At this point we should probably make some contact
with ongoing experiments. Can we test in some
simple way that the color structure due to the
non abelian theory is indeed correct? We have to
take into account fig. 5 and
recall some simple group relations and definitions
$$ \sum_{a\beta} T^a_{\alpha\beta} T^a_{\beta\gamma}
=C_F \delta_{\alpha\gamma},\qquad
\sum_{ac} T^a_{bc}T^a_{cd}= C_A \delta_{bd},\qquad
{\rm Tr}(T^a T^b)=T\delta_{ab}.\eqno(28)$$
In QCD $C_F=4/3$, $C_A=3$ and for generators
in the fundamental representation $T=T_F=1/2$, while
in the adjoint representation $T=T_A=3$. The
factors $C_F,C_A,T_F$ and $T_A$ appear in
jet counting rules, as shown in fig. 8

\vbox{\bigskip
\vskip 3 truecm
\medskip
\centerline{Fig. 8.- Lowest order jet cross sections.}
\bigskip}

\noindent
As evidenced by fig. 9 the experimental agreement
between the LEP data and
the QCD predictions is perfect.

\vbox{\bigskip
\vskip 5.5 truecm
\medskip
\centerline{Fig. 9.- QCD color factors as measured by ALEPH and
DELPHI. From$^{12)}$}
\bigskip}

\line{\bf 4.- Renormalization\hfil}
\medskip\noindent
Beyond tree level most Feynman diagrams
are ultraviolet divergent. Take for instance the
first diagram of those contributing at one loop
to the gluon propagator (fig. 10)

\vbox{\bigskip
\vskip 2.5truecm
\medskip
\centerline{Fig. 10.- One-loop contributions to the gluon self-energy.}
\bigskip}

\noindent
Neglecting external momenta, the integral over the
momenta of the internal particles is of the form
$$\int {{d^4k}\over{(2\pi)^4}}{{k^\alpha k^\beta}
\over k^4}=\infty\eqno(29)$$
To make sense of the theory  and get a finite result we must introduce
a cut-off $\Lambda$ and counterterms.
A typical method is to perform a subtraction at some
$q^2=-\mu^2$. For instance, for the self-energy
of the gluon propagator
$$\Pi(q^2)-\Pi(-\mu^2)\equiv\Pi_R(q^2)={\rm finite}.\eqno(30)$$
\vbox{\bigskip
\vskip 1.5truecm
\medskip
\centerline{Fig. 11.- Adding counterterms to make Green functions finite.}
\bigskip}
\noindent
Alternatively we can make sense of the integrals
using dimensional regularization by continuing
the dimensionality from 4 to $n=4+2\epsilon,$
$$\int {{d^4k}\over{(2\pi)^4}}\to\int{{d^nk}\over{(2\pi)^n}},\eqno(31)$$
and  subtract just the poles in $1/\epsilon$
(minimal subtraction, $MS$) or also the $\gamma_E-\log 4\pi$
that always accompanies the singularity in
$1/\epsilon$ (improved minimal subtraction, $\overline{MS}$).
For instance, for the second
diagram in fig. 10, namely the quark contribution
to the gluon self-energy, one has the following
result after computing the integral in $n=4+2\epsilon$
dimensions
$$
\Pi(q^2)=-{\alpha_s\over {6\pi}}\delta_{ab}(
{1\over\epsilon}+\gamma_E+\log{{m^2}\over{ 4\pi\mu^2}}+\dots).
 \eqno(32)$$
Using the $MS$ and $\overline{MS}$ schemes one gets
$$\Pi_{MS}(q^2)=-{\alpha_s\over {6\pi}}\delta_{ab}(
\gamma_E+\log{{m^2}\over{ 4\pi\mu^2}}+\dots)
\qquad
\Pi_{\overline{MS}} (q^2)=-{\alpha_s\over {6\pi}}\delta_{ab}(
\log{{m^2}\over{\mu^2}}+\dots)\eqno(33)$$
Admittedly this looks totally adhoc. It would seem that
we can get rid of any divergent integral by just
adding the appropriate counterterms. Of course, since these
are sort of arbitrary, we would then be able to obtain any
result we want. This is not so, of course. A renormalizable
theory is one in which the necessary counterterms to all
orders in perturbation theory are generated by
redefining the fields and parametres of the original
lagrangian and nothing else, and this is highly non-trivial. QCD is such a
theory:
if we redefine
$$\eqalign{
 g^0 &=Z_{1 YM} Z_{3 YM}^{-3/2} g
  =\tilde{Z}_1\tilde{Z}_3^{-1}
Z_{3 YM}^{-1/2} g
 = Z_{1 F} Z_{3 YM}^{-1/2} Z_{2 F}^{-1} g
= Z_5^{1/2} Z_{3 YM}^{-1} g,\cr
W_\mu^0 &= Z_{3 YM}^{1/2} W_\mu,\cr
\varphi^0 &= \tilde{Z}_3^{1/2}\varphi,\cr
\psi^0 &= Z_{2 F}^{1/2} \psi,\cr
{\xi_0} &= Z_6 Z_{3 YM}^{-1} {\xi},\cr}\eqno(34)
$$
it is possible
by a suitable non-unique choice of the $Z$'s to
make QCD finite. Since all $Z$'s are of
the form $Z=1+\Delta Z$, with $\Delta Z$ beginning at
${\cal O}(g^2)$ one must replace the original ${\cal L}_{QCD}$
one started with
by ${\cal L}_{QCD}+\Delta {\cal L}_{QCD}$.
$\Delta {\cal L}_{QCD}$ contains the necessary counterterms that must be
added. Equivalently, we can work directly with
the QCD lagangian with all fields and parametres
replaced by the `bare' ones, defined through eq. 34.
Observables will
be obtained by the prescription ($\Lambda$ being the
ultraviolet cut-off)
$$
\Gamma(p_i,g,\mu)=\lim_{\Lambda\to\infty}
Z_{3 YM}^{-m/2} Z_{2 F}^{-k/2}
\Gamma_0(p_i,g_0,\Lambda).\eqno(35)$$
On the l.h.s. we have renormalized Green functions or
amplitudes expressed as a function of
renormalized parametres. On the r.h.s we have
bare Green functions or amplitudes expressed as a
function of bare parametres.
In dimensional regularization
$$\Gamma(p_i,g,\mu)=\lim_{\epsilon\to 0}
Z_{3 YM}^{-m/2} Z_{2 F}^{-k/2}
\Gamma_0(p_i,g_0,\epsilon).\eqno(36)$$
$\mu$ is the subtraction scale and
 $m$ and $k$ are the number of
gluon and quark external lines, respectively. The effect of counterterms
is to
replace the dependence on the cut-off ($\Lambda$, $\epsilon$,
...) by a dependence on $\mu$. Unlike in QED where
the natural scale is $m_e$, or the Electroweak theory where
the natural scale is $M_W^2$, there is really no preferred
way of choosing the counterterms. The only requirement
is the fulfillment
of the Ward identities$^{13)}$. In practice $MS$ and $\overline{MS}$
are the most useful, particularly the latter that seems
to lead to perturbative series with a faster convergence. In the
$MS$ scheme all renormalization constants are just
poles in $\epsilon$
$$Z= 1+{\alpha_s \over \pi}{a\over \epsilon}
+({\alpha_s\over \pi})^2 ({b\over \epsilon^2}+
{c\over \epsilon})+\dots=
1+{\alpha_s^0 \over \pi}\mu^{2\epsilon}{a\over \epsilon}
+\dots \eqno(37)$$
\bigskip
\line{\bf 5.- Renormalization-group Equations\hfil}
\medskip\noindent
Note that regulated quantities depend on a cut-off
($\Lambda$, $\epsilon$, ...) and that the renormalization
of fields and constants through eq. 34
 trades the
dependence on the cut-off by some scale $\mu$. Yet,
physics cannot depend on $\mu$ at all. If you change $\mu$
you must change at the same time the value of your renormalized
parametres to make up for the difference. A simple
way to encode this observation is the following. Let's
write
$$\Gamma(p_i,\alpha_s,\mu)=Z_\Gamma(\mu,\epsilon)
\Gamma_0(p_i,\alpha_s^0,\epsilon)\eqno(38)$$
$\Gamma_0$ is obviously independent of the
subtraction scale. Therefore
$$ 0=\mu {d\over {d\mu}}\Gamma_0=
Z_\Gamma^{-1}( \mu {d\over {d\mu}}
-Z_\Gamma^{-1}\mu{d\over {d\mu}}Z_\Gamma) \Gamma.\eqno(39)$$
(We neglect everywhere the dependence on the gauge parameter,
as well as the quark masses. Of
course they have to be properly taken into account. Physical
on-shell amplitudes are $\xi$ independent.) From eq. 39
$$(\mu{\partial\over{\partial\mu}}+
\mu{{d\alpha_s}\over {d\mu}}{\partial\over{\partial\alpha_s}}
-Z^{-1}_\Gamma \mu{d\over {d\mu}}Z_\Gamma^{-1})\Gamma=0,\eqno(40)$$
or, defining the so-called $\beta$-function and
the Green function anomalous dimension $\gamma_\Gamma$,
$$\mu {{d \alpha_s}\over {d\mu}} = \alpha_s \beta(\alpha_s)
\qquad Z_\Gamma^{-1}\mu {d\over {d\mu}} Z_\Gamma^{-1}=\gamma_\Gamma,
\eqno(41)$$
$$(\mu{\partial\over {\partial \mu}}+
\beta\alpha_s {\partial\over {\partial \alpha_s}}-\gamma_\Gamma)
\Gamma(p_i,\alpha_s,\mu)=0.\eqno(42)$$
This is called the renormalization-group equation$^{14)}$. Let's
investigate its consequences. We can always write $\Gamma$
as a function of dimensionless variables by pulling out $\mu^D$
($D$: dimensionality of $\Gamma$)
$$\Gamma(\lambda p_i,\alpha_s,\mu)= \mu^D F({{\lambda^2 p_ip_j}
\over \mu^2}, \alpha_s).\eqno(43)$$
Hence
$$(\lambda {\partial\over {\partial \lambda}}
+\mu{\partial\over{\partial\mu}}-D)\Gamma (\lambda p_i,
\alpha_s,\mu)=0.\eqno(44)$$
Using the renormalization-group equation
we get
$$(-{\partial\over {\partial t}}+\beta\alpha_s{\partial\over
{\partial\alpha_s}}-\gamma_\Gamma +D)\Gamma(e^tp_i,\alpha_s,\mu)=0.
\eqno(45)$$
{}From subtraction scale independence arguments we have been
able to establish an equation concerning the dependence on the
external momenta. We
can formally solve this equation
$$\Gamma(e^tp_i,\alpha_s(\mu),\mu)=
\exp[tD-\int_0^t dt \gamma_\Gamma(\bar{\alpha}_s(t))]\times
\Gamma(p_i,\bar{\alpha}_s(t),\mu),\eqno(46)$$
 $\bar{\alpha}_s(t)$ is
just $\alpha_s(e^t\mu)$, i.e. the same coupling constant
but renormalized at a different scale.

{}From eq. 46 we see that when we scale
the external momenta in a Green function or amplitude the change is absorbed

\item{(i)}{in a multiplicative factor that depends on the
 anomalous dimension as well as the engineering dimension of the amplitude
and}
\item{(ii)}{in a redefinition of the coupling
$$\alpha_s(\mu)\to \alpha_s(e^t\mu).\eqno(47)$$}

\noindent
The renormalization-group evolution of the parametres in the
theory (in this case exemplified by the coupling constant
$\alpha_s$) governs the scaling behaviour.
We have to find which is the evolution of $\alpha_s$
under a change of $\mu$. This is actually a very simple
question. We just have to solve the equation
$$\mu {{d\alpha_s}\over {d\mu}}=\alpha_s\beta(\alpha_s).\eqno(48)$$
Since the $\mu$ dependence of $\alpha_s$ is introduced via counterterms,
To compute $\beta$ we have just to find the relevant renormalization
constants (see eq. 34).

Of course $\beta$ is evaluated in perturbation theory and,
accordingly, the differential equation eq. 48 is also
 solved in perturbation theory. The solution will
only make sense as long as the expansion parametre is
small. At one loop the solution of eq. 48 is
$$ \alpha_s(e^t\mu)={{\alpha_s(\mu)}
\over {1-{1\over \pi}\beta_1\alpha_s(\mu)t}}.\eqno(49)$$
In QCD $\beta_1$ (the first coefficient of the $\beta$-
function) is
$$\beta_1=-{11\over 2}+{N_f\over 3}\eqno(50)$$
(recall that in QED $\beta_1=2/3$). Note
that $\beta_1$ is negative if $N_f<16$. If $\beta_1$
is negative, at larger momentum transfers, where
the relevant scale will be $e^t\mu$ and not $\mu$,
$\alpha_s$ will actually decrease. The
solution of the renormalization-group equation
will actually become better and better at higher
energies.

\bigskip
\line{\bf 6.- Asymptotic Freedom\hfil}
\medskip\noindent
Whenever we speak of `higher' energies we are implicitly
assuming the existence of some characteristic QCD scale.
Looking at eq. 48
 we note that
$$t={1\over 2}\log \mu^2=\int{{d\alpha_s}\over
{\alpha_s \beta(\alpha_s)}}=\psi(\alpha_s)+C.
\eqno(51)$$
$C$ is an integration constant. Therefore
$t-\psi(\alpha_s)$ is a constant of motion along the
renormalization-group trajectory. At one loop,
we plug $\beta_1$ in eq. 51 and get
$${1\over 2}\log \mu^2+ {\pi\over {\beta_1\alpha_s(\mu)}}
=C\equiv {1\over 2}\log \Lambda^2_{QCD}\Rightarrow
\alpha_s(\mu)={-{\pi}\over{{\beta_1\over_2}
  \log(\mu^2/\Lambda_{QCD}^2)}}.\eqno(52)$$
If we renormalize at scales much larger than
$\Lambda_{QCD}$ the renormalized coupling
constant will be small and working at one loop
will be justified. Table 1 shows a number
of values for $\alpha_s$ extracted from different
LEP observables. The relevant scale here
is $M_Z^2$,
much larger than any hadronic scale. At such
energies perturbation theory is clearly meaningful.

\vbox{\bigskip\vskip 7.5truecm\medskip
\centerline{Table 1.- Determinations of $\alpha_s(M_Z)$.
{}From$^{12)}$.}\bigskip}

Let us recall the physical contents of eq. 46.
We can write it symbolically as

\vbox{\bigskip\vskip 3truecm\medskip
\centerline{Fig. 12.- Asymptotic freedom visualized.}
\bigskip}

\noindent
Ignoring for a second the overall factor, when
we scale up the momenta we have exactly the same
amplitude, but renormalized at scale
$e^t\mu$, i.e. replacing
$\alpha_s(\mu)$ by $\alpha_s(e^t\mu)$. Obviously
the scaled up amplitude will correspond to
a theory that interacts more weakly. In the
$t\to\infty$  limit
we will have a free theory. This is called
asymptotic freedom and it is one of the
characteristic signals of strong interactions.
QCD has it, while QED has not.

{}From fig. 12 we learn something else. Let us imagine that
for some reason radiative corrections are small for some
amplitude with momenta $p_i$ and the coupling constant renormalized
at scale $\mu$. (For instance, in QED we may decide to choose the
counterterms and {\it define} the renormalized coupling constant in
such a way that the classical formula for Thompson scattering
is strictly valid at all orders in perturbation theory for on-shell
particles.) The renormalization group  tells us is that if
we scale the external momenta {\it and} the renormalization scale
in the way prescribed by eq. 52,
radiative corrections will also be small for the scaled amplitude. Each
physical process has a characteristic scale. Choosing this scale as
subtraction point typically optimizes
the perturbative series.

We expect
the coupling constant $\alpha_s$
to be small at LEP energies
because $M_Z \gg \Lambda_{QCD}$, but
which evidence do we have that it actually runs
according to eq. 49?  Well, actually
$\alpha_s$ is not small enough for two-loop
effects to be completely neglected
so to answer this question in a precise way
we have to work a bit harder and compute
the two-loop $\beta$-function. The
result is
$$\beta_2=-{51\over 4} + {19\over 12} N_f.\eqno(53)$$
Incidentally, both $\beta_1$ and $\beta_2$ are
scheme independent (but not $\beta_3$ and beyond).
Note that, for $N_f<8$, $\beta_2$ is also negative.
Repeating the steps that led to eq. 52 we get
at the two-loop level
$$
{1\over 2}\log\mu^2+{\pi\over{\beta_1\alpha_s(\mu)}}
+{\beta_2\over \beta_1^2}\log
{\alpha_s(\mu)}+{\cal O}(\alpha_s)=
{1\over 2}\log \Lambda^2_{QCD},\eqno(54)$$
whose solution is
$$\alpha_s(\mu)={12\pi\over {(33-2N_f)\log(\mu^2/
\Lambda_{QCD}^2)}}\left[1-3{{153-19N_f}\over {(33-2N_f)^2}}
{{\log\log(\mu^2/\Lambda_{QCD}^2)}\over
{{1\over 2}\log(\mu^2/\Lambda_{QCD}^2)}}\right]
\eqno(55)$$
Working in the $MS$ or $\overline{MS}$ scheme
changes the value of $\alpha_s(\mu)$
but the difference is  ${\cal O}(\alpha_s^2)$.
Therefore the numerical value of $\Lambda_{QCD}$
does actually depend on the renormalization scheme
one is using (one does not see
this at the one-loop level precision). For instance,
at two loops $\Lambda_{MS}$ and
$\Lambda_{\overline{MS}}$ are related through
$$\Lambda_{MS}^2={e^{\gamma_E}\over {4\pi}}\Lambda^2_{\overline{MS}}.
\eqno(56)$$
Eq. 55 is thus our starting point to check
whether the behaviour predicted by the renormalization group
 has factual support.
However, there is one more thing that we
have to account for when we choose to
work in the $MS$ or $\overline{MS}$ schemes.
Decoupling of heavy fermions is not
manifest in any of these schemes. In practice
one works with the number of quarks that
are excited at the energy one is. For instance,
at scales well below and well above the
charm threshold we have, respectively (at the one-loop
level)
$$ \alpha_s(\mu)={{12\pi}\over
{27\log(\mu^2/\Lambda^2(3))}}
\qquad
\alpha_s(\mu)={{12\pi}\over
{25\log(\mu^2/\Lambda^2(4))}}.
\eqno(57)$$
The coupling constant is obviously continuous
as we cross the threshold, but $\Lambda_{QCD}$
is not. Demanding continuity on $\alpha_s$
leads to the matching condition
$$\Lambda(4)=\left({{\Lambda(3)}\over m_c}\right)^{2\over 25}
\Lambda(3).\eqno(58)$$
At two-loops this gets modified to$^{15)}$
$$
\Lambda(4)=\left({{\Lambda(3)}\over m_c}\right)^{2\over 25}
\left(\log{{m_c^2}\over {\Lambda^2(3)}}\right)^{-{{107}\over {1875}}}
\Lambda(3).\eqno(59)$$
If one insists in keeping $\Lambda(3)$ beyond the
charm threshold the corrections will be large;
perturbation theory breaks down.

The two loop evolution of the QCD coupling
constant is shown in fig. 13 for different
values of $\Lambda_{\overline{MS}}$. On top,
the values obtained from
different experiments at different scales
are shown. Asymptotic freedom is
very nicely confirmed. We cannot go
to energies much lower than those shown
because the coupling constant grows
very quickly and three-loop effects and beyond
are relevant; in fact, perturbation theory becomes meaningless. In the
coming sections
we will discuss some of the ways of determining $\alpha_s$ that are shown
in this plot.

\vbox{\bigskip\vskip 6.5truecm\medskip
\centerline{Fig. 13.- Evidence for next-to-leading scaling. From$^{16)}$}
\bigskip}

\line{\bf 7.- Confinement\hfil}
\medskip\noindent
$\Lambda_{QCD}$ sets a natural scale in
the theory. Well above $\Lambda_{QCD}$ perturbation
theory makes sense.  Of course perturbative QCD at
large enough energies describes a world
of quasi-free quarks, interacting with
Coulomb-like forces. We know very well
that hadronic physics is a very different
world with quarks are confined into
colorless hadrons.
As soon as
$q^2\sim\Lambda_{QCD}^2$ perturbation theory is
unreliable. It simply cannot explain confinement.

What does confinement actually mean? One
popular interpretation is that `it is
not possible to detect an isolated quark
or gluon'. The problem with this
definition of confinement is that
in electromagnetism, which is certainly
not confining, it is not possible to
detect an isolated electron either.
Electromagnetism (as well
as Quantum Chromodynamics) is long-range
(mediated by a massless particle) and plagued
(just as QCD) with infrared divergences.
Both QED and QCD observables have to be
inclusive enough. There is one difference, of course,
and that is that photons have
no $U(1)$ charge, while gluons
do carry $SU(3)$ charges. Far away from the
interaction point you can hope to be able to measure
the electric charge carried by one particle (or rather
what the experimental resolution defines as a particle
which is actually the electron surrounded by a cloud of
soft photons),
but even if you were able to construct a detector that
measured color you probably would not
be able to identify in any way the color of the quark itself.

In fact, there is another definition
of confinement that tells to you that the
chances of actually seeing a (gluon-dressed)
quark are small: `there is a force
between quarks that does not decrease with distance'.
There is indeed phenomenological evidence
(which is supported by lattice analysis) that the
interquark potential in QCD is of the
form
$$V(r)\sim a\Lambda^2_{QCD}r -{b\over r} +\dots
\eqno(60)$$
The first term is a confining quark potential. The
constant $a$ has to be  $\sim 1$ because $\Lambda^2_{QCD}$
 is the only dimensional quantity at our disposal.
The Coulombic part is called the L\"uscher term and plays
a crucial role in heavy quark spectroscopy$^{17)}$.

We like this second definition of confinement better,
because the first one is far too imprecise. In order to see
that it is useful to recall a toy example suggested
by Georgi$^{18)}$. Imagine a world in which we tune
$\Lambda_{QCD}$ in such a way that
$\alpha_s(1 {\rm GeV})=1/137$. Since the coupling
constant is so small, perturbation theory works
wonderfully at such energies. The proton would
be a bound state of quarks (bound by
Coulomb-like forces that is) with mass roughly
$3m_q$. Its size would be dictated by the
Bohr radius, about 1000 times the size it has
in our world. The inhabitants of this world
would certainly not understand the first
definition of confinement, since the confinement
radius would be $\sim \Lambda^{-1}\sim 10^{21}
{\rm cm}$. They would see quarks as you see
electrons.

Even in our world the situation is somewhat
similar to that of the toy world for
very heavy
quarks. Indeed  $\alpha_s(m_t)$ is
small (say $\sim 0.1$). The Bohr radius
is $r_0\sim 10^{-2}$ fm, much
smaller than $\Lambda_{QCD}^{-1}$. The
coulombic part of the interquark potential
largely dominates. (At such short
distances the linearly rising potential
is not at work, the leading confinement
effects are $\sim r^3$, as discussed
by Leutwyler some time ago$^{19)}$, but they can
be safely neglected at first approximation.)

Bottom and charm are in a somewhat intermediate
position. $\alpha_s(m_b)$ is still
relatively small. The Bohr radius
is $10^{-1}$ fm, smaller but comparable
to $\Lambda_{QCD}^{-1}$. Spectroscopy
is basically perturbative, at least for the lowest levels, but some
non-perturbative effects are visible. Charm is really no-man's land. Both
perturbative and non-perturbative effects compete even for the
ground state $n=1$.
For light quarks the Bohr radius is several fm and the confining
potential is fully at work.

\vbox{\bigskip\vskip 3.5truecm\medskip
\centerline{Fig. 14.- The QCD string.}\bigskip}

The existence of a confining potential leads
to very large
multiplicities and jets. The situation is
visualized in fig. 14. One can imagine
a quark-antiquark being formed at the
primary vertex then moving apart. Part of
their kinetic energy is deposited in the interquark
potential as they move away. Very quickly
a separation $r_m$ is reached where the energy deposited
is enough to form a new quark-antiquark pair,
$$\Lambda^2_{QCD} r_m\simeq 2 m_q,\eqno(61)$$
at that moment the quark-antiquark `string'
breaks and the process is repeated until
the average relative momentum is small
enough and hadronization takes
place.

There is a lot of physics in the string picture. We can think
of color forces being confined in some sort of tube or string joining
the two moving quarks. The chromodynamic energy is thus stored in
a relatively small region of space-time. If this picture is correct
we should expect hadronization to take place in this region in
preference to any other. This is indeed the case; in three jet events
(which originate from $\bar{q} q g$, with a hard gluon) there is a clear
enhancement
of soft gluon and hadron production in the regions between color
lines (representing the gluon by a double color line, or $\bar{q} q$
state), and a relative depletion in other regions.
This phenomenon is called color coherence$^{20)}$.

\bigskip
\line{\bf 8.- $R_{had}$\hfil} \medskip\noindent
The observable
$$R_{had}={{\sigma(e^+e^-\to hadrons)}\over {\sigma(e^+e^-\to
\mu^+\mu^-)}}\eqno(62)$$
is probably the cleanest and simplest
observable in QCD. It is fully inclusive and
can be computed through a dispersion relation which
is symbollically depicted in fig. 15

\vbox{\bigskip\vskip 3truecm\medskip
\centerline{Fig 15.- The imaginary part of the self-energy
describes the  $\gamma^*\to hadrons$ width.}\bigskip}

The sum over intermediate states on the r.h.s. of
fig. 15 runs over all hadronic states. We can just
as well compute this sum using another resolution of the identity---the one
that is provided to us by perturbative QCD, i.e. in
terms of quarks and gluons. $R_{had}$ has been
computed in this way up to third order in $\alpha_s$\footnote{$^1$} {We
shall write from now on $\alpha_s(\mu)$ or $\alpha_s(\mu^2)$ equivalently.
We have already argued that the right renormalization scale should be the
typical momentum transfer of the process.}
 $$R_{had}=R_0 [1+{{\alpha_s(q^2)}\over \pi}
+r_2({{\alpha_s(q^2)}\over \pi})^2+
r_3 ({{\alpha_s(q^2)}\over \pi})^3+\dots],
\eqno(63)$$
where $q^2$ is the typical momentum transfer, and in the $\overline{MS}$
scheme
$$r_2\simeq 2.0-0.12 N_f\qquad
r_3=-6.637 -1.200 N_f-0.005 N_f^2-1.240 {{(\sum Q_i)^2}
\over {3\sum Q_i^2}}.\eqno(64)$$
For $N_f=5$, $r_2\simeq 1.4$ and $r_3\simeq -12.8$.
We have seen that $\alpha_s(M_Z)\simeq 0.12$. Then
$$R_{had}=R_0[1+0.04+0.002-0.0008+\dots]
\eqno(65)$$
The convergence of the series does not look bad, but it is not very good
either. The values for the coefficients that we have been
just quoted correspond to massless quark; they
have to be accordingly modified for heavy quarks. When
this is done and comparison with experiment is
done we obtain from all four LEP experiments
$$\alpha_s(M_Z)=0.123\pm 0.008.\eqno(66)$$
The evolution of $\sigma(e^+ e^-\to hadrons)$ as a function of the energy
is visualized in fig. 16

\vbox{\bigskip\vskip 8truecm\medskip
\centerline{Fig. 16.- $\sigma(e^+ e^-\to hadrons)$. From$^{21)}$}
\bigskip}

Some comments are in order here. The first one is that
the convergence of the series and, ultimately, our
ability to get physics out of it hinges on two points.
 First of all, $\alpha_s$ has to be small
enough. A coupling constant $\alpha_s$ just slightly larger
produces a contribution for the ${\cal O}(\alpha_s^3)$
term comparable to the one of ${\cal O}(\alpha_s^2)$,
so the convergence of the series deteriorates
very quickly at low energies. Secondly, the
coefficients are actually scheme dependent. For instance,
if we repeat the calculation but in the $MS$ scheme
we get
$$r_2^{MS}=r_2^{\overline{MS}}+
(\log 4\pi - \gamma_E){{33-2N_f}\over 12}
\simeq 7.35-0.44 N_f.\eqno(67)$$
The convergence of the perturbative series is
much worse in the $MS$ scheme. It is not clear
why it is so good in the $\overline{MS}$, because
in Field Theory this is often the case  only if one
uses a physically motivated scheme (such
a subtraction at some energy scale), while
the reasons to use the $\overline{MS}$ scheme
are of practical order. Be as it may, this
is a welcome fact.

$R_{had}$ is an observable so it must
be independent of the renormalization scheme.
If we work in the $MS$ scheme something else
must change so that the net result is still
the same. The thing that changes is of course
the coupling constant itself
$$\alpha_s^{\overline{MS}}(\mu)-\alpha_s^{MS}(\mu)
= {{12\pi}\over {(33-2N_f)\log{(\mu^2/ \Lambda_{\overline{MS}}^2)}}}
{{\log 4\pi-\gamma_E}\over {\log{(\mu^2/\Lambda_{\overline{MS}}^2)}}},
\eqno(68)$$
which just makes up
for the changes (up to $\log\log$ terms), i.e.
$$1+{{\alpha_s^{MS}(\mu)}\over \pi}+r_2^{MS}
({{\alpha_s^{MS}(\mu)}\over \pi})^2+\dots =
1+{{\alpha_s^{\overline{MS}}(\mu)}\over \pi}+r_2^{\overline{MS}}
({{\alpha_s^{\overline{MS}}(\mu)}\over \pi})^2+\dots
\eqno(69)$$
This is a general feature of perturbation
theory. In this framework we
necessarily deal with
truncated series, so independence of the
subtraction point or of the scheme can only be checked up to
terms of the next order in the expansion. That is
($R$ denotes some renormalization prescription)
$$\sum_{n=0}^\infty c_n(R)\alpha_s(R)^n=
\sum_{n=0}^\infty c_n(R^\prime)\alpha_s(R^\prime)^n,
\eqno(70)$$
but
$$
\sum_{n=0}^N c_n(R)\alpha_s(R)^n\neq
\sum_{n=0}^N c_n(R^\prime)\alpha_s(R^\prime)^n;
\eqno(71)$$
the error being of ${\cal O}(\alpha_s^{N+1})$.

A lot of theoretical work has been done on the
issue of which is the `optimal' scheme$^{22)}$. Some
possibilities are

\item{(i)}{ FAC (Fastest Apparent Convergence):
choose the scheme that makes $c_N=0$, $c_N$ being
the last computed coefficient.}
\item{(ii)}{ PMS (Principle of Minimal Sensitivity):
demand that
$${{\partial}\over {\partial R}}(observable)=0,$$
and work in the scheme $R$ that fulfills this equation
(which, of course, would be {\it any} scheme if we knew the
observable exactly).}

It is a fact that the quality of the series improves
when one uses these methods, but unfortunately one is
forced in general to use different schemes for different
observables. On the basis of these analysis it has even
been claimed that $\alpha_s$ somehow `freezes' at
$\sim 0.3$ at low energies. However, it is fair to say
that the general properties of the perturbative series in QCD are so poorly
understood that any method that does not directly
rely on actually computing the neglected first term
in the perturbative expansion
and making sure that it is small
is likely to be met with
skepticism. It is hard to base derivations of
fundamental parametres such as $\alpha_s$ on
`optimization'
techniques.
\bigskip\bigskip
\line{\bf 9.- $R_{\tau}$\hfil}
\medskip\noindent
On energy considerations it is obvious
that the $\tau$ is the only lepton we know
heavy enough to decay into hadrons. This
of course makes it a very interesting object.
The inclusive decay rate $\tau\to \nu_\tau+
hadrons$ can, in principle, be derived from QCD
by exactly the same techniques as $R_{had}$.
This is shown in fig. 17.

\vbox{\bigskip\vskip 3.5 truecm\medskip
\centerline{Fig. 17.- Determination of $R_\tau$ through
dispersion relations.}\bigskip}

The counterpart of $R_{had}$ is now
$$R_\tau={{\Gamma(\tau\to \nu_\tau+ hadrons)}
\over {\Gamma(\tau\to \nu_\tau e \bar{\nu}_e)}}.
\eqno(72)$$
At lowest order in QCD
$$R_\tau={{\Gamma(\tau\to \nu_\tau d \bar{u})+
\Gamma(\tau\to \nu_\tau s \bar{u})}
\over {\Gamma(\tau\to \nu_\tau e \bar{\nu}_e)}}
\simeq 3.
\eqno(73)$$
$R_\tau$ can be computed as a power series in
$\alpha_s$. Unlike in $R_{had}$, the typical scale will be
very low (even zero) due to kinematical reasons
(see fig. 17). To tackle this problem we
decompose the $W$ boson self-energy
(which here plays the role  the photon
vacuum polarization did in $R_{had}$) into
vector and axial parts as well as
Cabibbo-allowed and Cabibbo-suppressed
terms
$$ \Pi^{\mu\nu}= \vert V_{ud}\vert^2
(\Pi^{\mu\nu}_{ud V}+\Pi^{\mu\nu}_{ud A})
+\vert V_{us}\vert^2 (\Pi^{\mu\nu}_{us V}+\Pi^{\mu\nu}_{us A}).
\eqno(74)$$
Each $\Pi^{\mu\nu}$ can be split into transverse and
longitudinal parts
$$
\Pi^{\mu\nu}= (-g^{\mu\nu}k^2 +k^\mu k^\nu)\Pi^{(1)}
+k^\mu k^\nu \Pi^{(0)}.\eqno(75)$$
For massless fermions $\Pi^{(0)}=0$. We assume that this is the case for
simplicity. For a detailed account see$^{23)}$. We can write
$$\eqalign{
R_\tau & =12\pi\int_0^{m_\tau^2}
{{ds}\over {m_\tau^2}} (1-{s\over m_\tau^2})^2
(1+{{2s}\over m_\tau^2}) {\rm Im} \Pi^{(1)}(s)\cr
& =  6\pi i \int_{\vert s\vert=m_\tau^2}
{{ds}\over {m_\tau^2}} (1-{s\over m_\tau^2})^2
(1+{{2s}\over m_\tau^2})  \Pi^{(1)}(s)\cr}
\eqno(76)$$
where we have used Cauchy theorem to write the
integral we are interested in as a contour
integral. For large $\vert s\vert$ we can compute
$\Pi^{(1)}(s)$ just as we did for
the hadronic vacuum polarization. One
gets basically the same result (up to factors). For
instance
$$ {\rm Im}\Pi^{(1)}(s)=
{1\over {2\pi}} (\vert V_{ud}\vert^2 +
\vert V_{us}\vert^2)[ 1+ {{\alpha_s(s)} \over \pi}
+ r_2 ({{\alpha_s(s)} \over \pi})^2+ r_3({{\alpha_s(s)} \over \pi})^3
+\dots].\eqno(77)$$
We also know that
$$ {{\alpha_s(s)}\over \pi}=
{{\alpha_s(m_\tau^2)}\over \pi}+{\beta_1\over 2}
{{\alpha_s(m_\tau^2)}\over \pi}\log{ s\over
m_\tau^2}+\dots
\eqno(78)$$
So, finally
$$ R_{\tau}=
3(\vert V_{ud}\vert^2 +\vert V_{us}\vert^2)
[ 1+ {{\alpha_s(m_\tau^2)} \over \pi}
+ (r_2-{19\over {24}}\beta_1) ({{\alpha_s(m_\tau^2)} \over \pi})^2+
\dots + {\rm n.p.t.}]
\eqno(79)
$$
n.p.t. stands for non-perturbative contributions
(contributions non-expressable as a power series in $\alpha_s$).
We did not have to worry too much about them in $R_{had}$
because there
they were characteristically suppressed by a
power of the momentum transfer. Here they
can be important. Fortunately, they have been
found to be small$^{23)}$. Anyhow, we are now in
posession of an alternative way of
determining $\alpha_s$ though $\tau$ decay.
The best data comes again from LEP. A fit
to experimental numbers (including mass
corrections, which have been neglected throughout)
gives
$$ \alpha_s(m_\tau)= 0.341\pm 0.035 \qquad\Rightarrow\qquad
\alpha_s(M_Z)= 0.121\pm 0.004.
\eqno(80)$$
Thanks to the logarithmic scaling this
method provides us with the most accurate
determination of $\alpha_s$ so far. It is
also a nice way of testing next to leading scaling.
\bigskip
\line{\bf 10.- Logs in QCD\hfil}
\medskip\noindent
Before continuing our discussion, it is convenient
to make a short theoretical digression. We
have seen how the scaling  of $\alpha_s$ with the energy
is logarithmic. Logs in fact play a very crucial
role in Quantum Field Theory. Let us then
stop and think for a second which is the
origin of these logarithmic terms.

Actually there are two types of logarithms in a Field
Theory such as QCD
$$ \log{q^2\over \mu^2}\qquad \log{q^2\over \lambda^2}
\eqno(81)$$
They have very different origin. $\mu$ is
some renormalization or subtraction scale,
while $\lambda^2$ can be some external momentum squared or a small
(mass)$^2$ that we have given by hand to the gluon. The first type are
associated to ultraviolet divergent integrals (integrals with a bad behaviour
when the internal momentum is large). The second type
are infrared logs
 and are related to Feynman diagrams
with a bad behaviour when one or more external
momenta vanish.
Ultraviolet logs
appear in any renormalizable Field Theory after
renormalization. On the contrary, infrared logs
appear whenever a theory has massless particles in
the spectrum (such as photons or gluons).

A given Feyman diagram can give rise to both
type of singularities at the same time. This is
illustrated in fig. 18

\vbox{\bigskip\vskip 3truecm \medskip
\centerline{Fig. 18.- Infrared logs.} \bigskip}

There actually two classes of infrared logs caused
by massless particles. The so-called
infrared divergences arise from the
presence of a {\it soft massless} particle
($k^\mu\to 0$). For instance in the
process $e^+e^-\to \mu^+\mu^-$ at the
one loop level (fig. 19)
we have to compute the integral

\vbox{\bigskip\vskip 3truecm\medskip
\centerline{Fig. 19.- Example of diagram having an infrared divergence.}
\bigskip}

$$\int {{d^4k}\over {(2\pi)^4}}
{1\over { k^2 [(p_1+k)^2-m^2][(p_2+k)^2-m^2]}}.
\eqno(82)$$
When $p_1^2=p_2^2=m^2$ the integral
behaves for $k^\mu\to 0$ as
$$\int{{d^4k}\over {(2\pi)^4}}{1\over k^4}.\eqno(83)$$
and diverges. This divergence is
unphysical so it must be cancelled by something
else. The Bloch-Nordsieck theorem$^{24)}$ states that
in inclusive enough cross-sections the
infrared logs cancel. What do we mean by `inclusive
enough'?  A detector will not be able to discern
a `true' muon from a muon accompanied by a soft
enough photon (with $\vec{k}\to 0$). Therefore,
in addition of the diagram shown in fig. 20 we have
to consider diagrams where a soft photon is
radiated by the muon, square the modulus of the
amplitude and integrate over the available phase
space (which actually depends on the experimental cut). When
this is done the result is infrared finite. The
relevant diagrams are depicted in fig. 20

\vbox{\bigskip\vskip 3truecm\medskip
\centerline{Fig. 20.- Real and virtual photons have to be included for IR
safe results.} \bigskip}

The other type of infrared logs are called mass singularities.
They occur in theories with massless particles
because two {\it parallel massless} particles have an
invariant mass equal to zero
$$ k^2=(k_1+k_2)^2=\Vert (\omega_1+\omega_2,0,0,
\omega_1+\omega_2)\Vert=0.\eqno(84)$$
The appeareance of such a mass singularity is
illustrated in fig. 21

\vbox{\bigskip\vskip 3truecm\medskip
\centerline{Fig. 21.- Diagram with a mass singularity.}\bigskip}

$${1\over {(p-k)^2}}={1\over {p^2+k^2-2k^0p^0+
2k^0p^0\cos \theta}},\eqno(85)$$
the denominator vanishes when we set all particles
on shell ($p^2=k^2=0$) and $\theta\to 0$
(i.e. $\vec{k}$ is parallel to $\vec{p}$). Even if one of the two particles
is massive there is a singularity, provided the
3-momenta are parallel.

The Kinoshita-Lee-Nauenberg theorem$^{25)}$ ensures
that for inclusive enough cros section the mass
singularities also cancel. Both for mass singularities
and for infrared divergences there is a trade-off
between $\lambda^2$, the infrared regulator of a massless particle,
and the energy and angle resolution of the
inclusive cross section $\Delta E$, $\Delta \theta$.

In practice, it is better to regulate the infrared logs using
dimensional regularization (introducing $\lambda^2$ leads to
difficulties with gauge invariance). Real gluon emission
diagrams are regulated by performing the phase space integration
in $n$ dimensions.

There is in fact a lot of physical insight
hidden in the infrared logs. We have seen that the
contribution from the diagram in fig. 19 is infrared
divergent, i.e. {\it infinite}. Yet, physical arguments
tell us that the probability of finding a `bare' isolated
muon should be {\it zero}. We know this because
detectors are unable to tell apart a muon from
a muon plus one soft photon or indeed from a
muon plus any number of soft photons.
Infrared divergences in QED can be summed up and
then one sees that the probability of finding
an isolated muon is indeed zero and not infinite as
the one loop diagram led us to believe. Whenever
a Feynman diagram is infrared divergent it means
that we have forgotten something relevant.

Let us consider in QED the interaction of a charged
fermion with an external source  and let us expand
in the number of {\it virtual} photons $n$
(i.e. in the number of loops). The total
amplitude will be expressed
as
$$ M(p,p^\prime)=\sum_{n=0}^\infty M_n(p,p^\prime)
\eqno(86)$$
then a calculation shows that
$$\eqalign{M_0 &=m_0, \cr
	   M_1 &=m_0\alpha B+m_1,\cr
	   M_2 &=m_0{{(\alpha B)^2}\over 2}+m_1\alpha B + m_2,\cr
	       &\dots \cr}\eqno(87)$$
The quantities $m_n$ are IR-finite, while $B$ is IR-divergent.
The series in eq. 86 can be summed up
$$M=\exp(\alpha B)\sum_{n=0}^\infty m_n,\qquad m_n\sim\alpha^n,
\eqno(88)$$
and
$B$ can be obtained just from the lowest order diagram. Introducing
an IR cut-off $\lambda$,
$B\sim -\log {m^2/ \lambda^2}$,
which indeed shows that when we remove the cut-off the probability
of finding an isolated charged fermion is zero
in QED. The addition
of soft photons changes that result multiplying the
total amplitude by a factor $\sim (\Delta E/\lambda)^2$.
There is a trade between the infrared regulator
and $\Delta E, \Delta \theta$. The latter are, of course, detector-
dependent quantities.

Although only partial results exist$^{26)}$, it is believed that
a similar exponentiation takes place in QCD. Due to the
confinement subtleties it is unclear whether the
suppression factor is compensated by radiation
of soft gluons. Even if this compensation does actually take
place that would not disprove confinement, only that
confinement would have nothing to do with the
structure of infrared singularities of the theory.
\bigskip
\line{\bf 11.- Jets and $\alpha_s$\hfil}
\medskip\noindent
The previous discussion can be summarized in the following way: due to IR
singularities one is forced to consider
cross sections not of individual
particles in the final state, but rather of
bunches of particles, each `hard' quark
and gluon surrounded by a `soft' cloud of
gluons and, perhaps, quarks.
We will call these bunches `jets'.

The Bloch-Nordsieck and Kinoshita-Lee-Nauenberg
theorems guarantee the finiteness of the
cross-sections. We have to define
an energy and angle resolution. For
instance, if $p$ is the momentum of a
primary quark (see fig. 22)
we can impose that the energy of each
soft particle in its jet satisfies
$k^0_i<\epsilon p_0$ and also that
$\arg (\vec{p},\vec{k}_i)<\delta$.
We will get singularities
of the form $\alpha_s\log\epsilon\log\delta$
when $\epsilon,\delta\to 0$. The specific details depend
on the precise definition of the jet.
Popular jet algorithms have been
discussed in this Meeting$^{2)}$.

\vbox{\bigskip\vskip 3.5 truecm\medskip
\centerline{Fig. 22.- Idealization of a jet.}
\bigskip}

The situation is unfortunately even more
involved because  hadrons and not
quarks and gluons are detected. The
evolution of the quarks and gluons produced at high
momentum
transfer is perturbative at first, until the average separation of
the particles becomes ${\cal O}(\Lambda_{QCD}^{-1})$. Then the confining
potential (and the string picture) takes over. Eventually
one is forced for anything other than fully
inclusive observables (such as $R_{had}$) to introduce fragmentation and
hadronization models to compute the
observable cross-sections.
Any observable that is not fully inclusive
is described by a convolution of two
very different types of physics
$$
{\rm Observable=Perturbative}\otimes
{\rm Non-perturbative}
$$
The perturbative part is, in principle,
calculable in QCD as a power series in
$\alpha_s$. It is affected by some
`theoretical' uncertainties since
most observables have been calculated up
to next to leading order, and not beyond, and
higher order effects can be important.
In addition there is the issue of the choice
of an adecuate renormalization scale, which sometimes
is far from obvious.

 The non-perturbative part has
to be modelled. Its relation to QCD
and its parametres (such as $\alpha_s$)
is unclear. A considerable amount
of cross-checking and experimental
feed-back is required.
In general, the more inclusive the
observable is the smaller the
unknowns coming in from the
non-perturbative part are. At LEP there are
a number of observables that are widely used
and where the dependence on
the hadronization model is believed to be
under control such as $R_3$,
thrust or energy-energy correlations.
The obvious question is do they give
consistent results? The question
is partially answered in
the positive in fig. 23, showing very
good agreement between determinations
of $\alpha_s$ from jet topologies.

A big improvement is obtained after the inclusion of
 next to leading corrections. They are clearly
required; the data would be just inconsistent otherwise.
Fig. 24 shows the energy evolution
of $R_3$, the ratio between three and two jets
which is directly proportional to $\alpha_s$
compared to the QCD two-loop prediction. From this
observable alone the precision on $\alpha_s$ is
approximately $\pm 0.010$, which is, by itself,
quite remarkable. Not too long ago, at PETRA
energies, the determination of $\alpha_s$
was painstakingly difficult. Why is it so much easier
at LEP energies than hitherto?

To understand this we look in some more detail
at the thrust of the jets produced.
Thrust is defined as
$$T={\rm max}{{\vert \sum_i \vec{p}_i \vec{n}_T\vert}
\over{\sum_i\vert\vec{p}_i\vert}}\eqno(89)$$
$\vec{n}_T$ is the thrust axis, which is varied
to maximize $T$. $T$ takes values between 0.5 (spherical)
to $T=1$ (complete alignment). Figure 25 shows the average
thrust for a number of experiments. The LEP
value is close to 0.94, higher than in any previous experiment; events
are well aligned with the momentum of the primary
quark. As a consequence, it is much easier
to count jets at LEP
than in any previous machine.

\vbox{\bigskip\vskip 8truecm\medskip
\noindent Fig. 23.- Determination of $\alpha_s(M_Z)$ from
different observables by OPAL$^{27)}$. The figure includes leading
and next to leading corrections and two different hadronization models.
Observables are (left to right): thrust, oblateness, F-Major, C-Par, $R_3$.
\bigskip}

\vbox{\bigskip\vskip 5.5truecm\medskip
\centerline{Fig. 24.- $R_3$ for different experiments.}
\bigskip}

At LEP energies we are in an energy
range where perturbative QCD calculations seem to exhibit
a reasonable convergence. If we consider
the average value $\alpha_s(\mu=34 {\rm \,GeV})=
0.158\pm 0.020$ obtained from $e^+ e^-$ machines, excluding LEP (but
including Tristan data), with very similar, if not identical,
hadronization models, jet algorithms, etc. we
see that an  30\% increase in the coupling constant
enlarges the experimental error by more that a factor 3. Radiative
corrections rapidly become hard to control.
The size of the error is crucial
 to be able to extract physics from the measurements of $\alpha_s$.
For instance, the previously quoted value of $\alpha_s$ leads
to $\Lambda_{\overline{MS}}=440^{+320}_{-220}$ MeV, while
from the LEP results one gets $\Lambda_{\overline{MS}}=250^{+100}_{-80}$
MeV, a much more stringent result. Incidentally, LEP is the first
experiment where the ${\cal O}(\alpha_s^2)$ perturbative contribution
is bigger than the error induced by hadronization models.

\vbox{\bigskip\vskip 5.5truecm\medskip
\centerline{Fig. 25.- Narrowing of jets with increasing $E_{cm}$.
} \bigskip}

\vbox{\bigskip \vskip 8.5truecm\medskip
\centerline{Table 2}
\bigskip}

\bigskip
\line{\bf 12.- D.I.S.: Free Parton Model\hfil}
\medskip\noindent
A brilliant confirmation
of the existence of nearly free constituents
inside the nucleon was provided more than
twenty years ago by a series of experiments
carried out at SLAC$^{28)}$. Then it became possible
to scatter electrons off nucleons in fixed
target experiments with a typical momentum transfer
$\sim 1 - 10$ (GeV)$^2$, a kinematical range unexplored until that time.
The
kinematics of Deep Inelastic Scattering processes
is shown in fig. 3.

The virtual intermediate boson
is far off its mass-shell and
scatters off a quark or gluon in a time
of ${\cal O}({1/{\sqrt{-q^2}}})$. Typically
quarks and gluons are themselves off-shell by an
amount of ${\cal O}(\Lambda_{QCD})$. After
the scattering the outgoing particles recombine
into hadrons in a time of ${\cal O}(1/\Lambda_{QCD})$.
Thus Deep Inelastic is a two-step process
\item{(i)}{ Short distance scattering occurs with a large
momentum transfer. Well described by perturbation theory.}
\item{(ii)}{ Outgoing particles recombine. Not calculable
in perturbation theory.}
However, step (ii) can be side-stepped all together
for fully inclusive rates. Then perturbation theory
is adecuate to describe many features of DIS.

If we place ourselves in the center of mass of the
hadron and virtual intermediate boson both particles
move very fast towards each other. Whatever components
the hadron contains they will all have
moments parallel to $P^\mu$, up to transversal
motion of ${\cal O}(\Lambda_{QCD})$. Let us write
$$p^\mu=x P^\mu.\eqno(90)$$
The squared CM energy of the lepton and proton constituent
will be
$$\hat{s}=(xP+k)^2\simeq 2xPk\simeq xs.\eqno(91)$$
We
neglect masses (as well as the fact that constituents are
off-shell
by ${\cal O}(\Lambda_{QCD})$)
The final momentum of the constituent is $xP+q$.
Therefore
$$0\simeq (xP+q)^2\simeq 2x Pq +q^2,\eqno(92)$$
so $x=-q^2/2Pq$. If $\nu$ is the energy
transfer in the LAB system, we can also
write
$$x={{-q^2}\over{2\nu m_N}}\eqno(93)$$
$m_N$ being the nucleon mass. It is convenient to
introduce
$$y={{Pq}\over {Pk}}=1-{{Pk^\prime}\over {P k}}\eqno(94)$$
In the lab frame $y=\nu/E$ and $0\le y\le 1$. $y$ is thus
the relative energy loss of the colliding lepton.

Let us for the time being ignore altogether QCD
interactions and let us assume that constituents
of the nucleons (which we will call partons) are
free. DIS will then be described
by an incoherent sum over elementary processes. The partonic
differential cross sections
in the LAB frame will be

\noindent
$\bullet$ $\nu q$, $\overline{\nu} q$-scattering

$${{d\sigma_\nu}\over {dy}}
=({{g^2}\over {4\pi}})^2{{\pi m E}\over {(q^2-M_W^2)^2}}
[g_L^2+g_R^2(1-y)^2], \eqno(95)$$

$${{d\sigma_{\overline{\nu}}}\over {dy}}
=({{g^2}\over {4\pi}})^2{{\pi m E}\over {(q^2-M_W^2)^2}}
[g_R^2+g_L^2(1-y)^2]. \eqno(96)$$

\noindent
$\bullet$ $eq$-scattering

$${{d\sigma_e}\over {dy}}
=Q^2 {{4\pi\alpha^2m E}\over {q^4}}[1+(1-y)^2].\eqno(97)$$
The neutral current sector is dominated by $\gamma$ interchange
below $q^2=M_Z^2$, so we have not bothered to include $Z$
exchange. In eq. 97 $Q$ is the quark electric charge (in units
of $e$) and $m$ is the
target mass. Since $p^\mu=x P^\mu$, we just
take $m=x m_N$. Then, for instance,
$${{d^2\sigma_e}\over {dx dy}}
=Q^2 {{4\pi\alpha^2 xm_N E}\over {q^4}}[1+(1-y)^2].\eqno(98)$$

Let $u(x)dx, d(x)dx,...$ be the number of $u,d,...$
quarks with momentum fraction between $x$ and $x+dx$
in a nucleon. Then $xu(x), xd(x),...$ will be
the fraction of the nucleon momentum carried by
$u,d,...$ quarks. We, of course, identify  quarks
with partons and, since we assume that they are free, proceed to
sum incoherently over the different scattering
possibilities. For instance in
$ep\to eX$
$$
{{d^2\sigma}\over {dxdy}}= {{2\pi\alpha^2}\over s}
{{1+(1-y)^2}\over {xy^2}}
[{4\over 9}(u(x)+\overline{u}(x))+
{1\over 9}(d(x)+\overline{d}(x))+
{1\over 9}(s(x)+\overline{s}(x))].\eqno(99)$$
(We neglect here the possible contribution from the
sea of heavy quarks in the nucleon.) Other
DIS processes weigh differently quarks
and antiquarks. For instance, in $\nu p\to \mu X$
if $-q^2\ll M_W^2$ we have
$$ {{d^2\sigma}\over {dxdy}}= x {{G_F^2 s}\over
\pi}[ c_c^2 d(x)+s_c^2 s(x)+\overline{u}(x)(1-y)^2],
\eqno(100)$$
with $c_c=\cos \theta_c$, $s_c=\sin \theta_c$, the
cosinus and sinus of the Cabibbo angle, respectively.

 The parton distribution functions $q(x)$ are
quantities which at present, generally speaking,
cannot be derived from
QCD. The other way round is probably more interesting: we can learn a lot
about the non-perturbative regime of QCD from DIS experiments through
these parton distribution functions (PDF). Probably the first thing
that one learns from them is that gluons are very important.
{}From the SLAC-MIT data$^{28)}$
$$Q=U+D+S=\int_0^1 dx x (u(x)+d(x)+s(x))\simeq 0.44,\eqno(101)$$
$$\bar{Q}=\bar{U}+\bar{D}+\bar{S}=
\int_0^1 dx x (\bar{u}(x)+\bar{d}(x)+\bar{s}(x))\simeq 0.07.
\eqno(102)$$
The total fraction of momentum carried by quarks (and
antiquarks) is only about 50\% The rest is carried by gluons,
showing that although the naive quark model works very
well is just a gross simplification as a model of hadrons, at least at
large $-q^2$.

Another example that the quark model fails to describe some
basic features of hadrons is provided by the `spin of the proton'
problem$^{3)}$. $\mu$-scattering on polarized targets shows that
the fraction of the total spin of the proton that can naively be
associated to constituent quarks is surprisingly small. We shall
not dwelve on this matter further here.

Nevertheless,
there are some obvious sum rules for the parton
distribution functions which can  ultimately be explained in terms
of the quark model. For the proton
$$\int_0^1 dx (u(x)-\bar{u}(x))= 2,\eqno(103)$$
$$\int_0^1 dx (d(x)-\bar{d}(x))= 1,\eqno(104)$$
$$\int_0^1 dx (s(x)-\bar{s}(x))= 0.\eqno(105)$$
On QCD grounds we expect that this free parton model
description of the hadrons becomes more and more accurate
when $-q^2\to \infty$, $\nu\to\infty$, while
keeping $x$ fixed. This limit is known as
Bjorken scaling and in the strict $-q^2=\infty$ limit everything depends
just on $x$.

The parton distribution functions are actually a function
of the energy. At $-q^2=\infty$ QCD is a free theory, so we
can imagine quarks and gluons sharing in equal terms
the total momentum, in a sort of QCD version of the
equipartition theorem of Statistical Mechanics. Therefore
we expect that in this limit the relation
between the momentum carried
by quarks and the one carried by
gluons should be  $N_cN_f/ 2(N_c^2-1)$. This can be
rigorously justified within QCD$^{29)}$. From the above
limiting value we see
 and at higher energies the total
momentum carried by {\it constituent} or {\it valence} quarks
diminishes and that an equally important role is played
by particles from the Dirac sea of the nucleon.

Let us now try to rederive the previous results in a
more theoretical setting. Let us consider
for instance $\nu p$ scattering. Then
$${{d^2\sigma}\over {d(-q^2) d\nu}}=
{{G_F^2 m_N}\over {\pi s^2}} L^{\mu\nu} H_{\mu\nu},
\eqno(106)$$
where
$$L^{\mu\nu}={1\over 8}{\rm Tr}
[\gamma^\mu(1-\gamma_5)\gamma^\alpha\gamma^\nu
(1-\gamma_5)\gamma^\beta] k_\alpha k_\beta \eqno(107)$$
is the trace over the leptonic external lines, and $H_{\mu\nu}$
is given by
$$
\sum_{X}\langle P\vert J_\mu(0)\vert X(P^\prime)\rangle
\langle X(P^\prime)\vert J_\nu(0)\vert P\rangle
=\int d^4z e^{iqz}\langle P\vert J_\mu(z)J_\nu(0)\vert
P\rangle
={\rm Im}\Pi_{\mu\nu}(q),\eqno(108)$$
with
$$\Pi_{\mu\nu}(q)=\int d^4z e^{iqz}
\langle P\vert T J_\mu(z)J_\nu(0)\vert P\rangle.\eqno(109)$$
We decompose $H_{\mu\nu}$ as
$$H_{\mu\nu}=-g_{\mu\nu}F_1+{{P_\mu P_\nu}\over {\nu m_N}}F_2
+{i\over {2\nu m_N}}\epsilon_{\mu\nu\rho\sigma}P^\rho q^\sigma F_3
\eqno(110)$$
(If we assume that we are working with non-polarized targets $P$ and
$q$ are the only vectors at our disposal.) $F_1$, $F_2$ and $F_3$ are
the nucleon structure functions.
Using the kinematical relations
$x=-q^2/2\nu m_N$ and $y=2m_N\nu/ s$ we get
$$ d(-q^2) d\nu= \nu s dx dy,\eqno(111)$$
$${{d^2\sigma}\over {dx dy}}={{G_F^2 s}\over {2\pi}}
[F_1 xy^2+ F_2(1-y)-F_3xy(1-{y\over 2})].\eqno(112)$$
Let us now compare with the free parton model. We see
that (restoring the $\nu p$ index, to make
apparent that the structure functions
are process dependent)
$$ F_1^{\nu p}(x)=c_c^2(\bar{u}(x)+d(x))+s^2_c(s(x)+\bar{u}(x)),$$
$$ F_2^{\nu p}(x)=
2xc_c^2(\bar{u}(x)+d(x))+2xs^2_c(s(x)+\bar{u}(x)),\eqno(113)$$
$$ F_3^{\nu p}(x)= 2c^2_c(\bar{u}(x)-d(x))
+2 s^2_c(-s(x)+\bar{u}(x)).$$
For other processes the actual expressions may vary but
the structure functions are always linear combinations of
the parton distribution functions, i.e. $F_2(x)=x\sum_i
\delta_i q_i(x)$, etc.
Note that in the free parton model
$$F_L(x)=F_2(x)-{{F_1(x)}\over {2x}}=0.\eqno(114)$$
This is the
Callan-Gross relation, which actually is not an  exact one;
it gets modified when the $q^2$ dependence is
included, i.e. we depart from the strict $-q^2=\infty$ limit.
$F_L$ in some sense measures the spin of the target.
Let us assume for one second that our target has
spin zero instead of one half.
 Then we can write
with the help of just one form factor
$$\langle X(P^\prime)\vert J_\mu(0)\vert P\rangle
= F(q^2) (p+p^\prime)_\mu \eqno(115)$$
Plugging this back into eq. 108 we immediately
observe that $F_1$=0. However, experimentally
the Callan-Gross relation is well satisfied, and,
on the other hand, $F_2$ is certainly non-zero
(see fig. 26 below).
\bigskip
\line{\bf 13.- Scaling Violations\hfil}
\medskip\noindent
Fig. 26 shows some recent data from the ZEUS
collaboration at HERA$^{30)}$ .

\vbox{\bigskip\vskip 9 truecm\medskip
\centerline{Fig. 26.- $F_2$ as measured by ZEUS.}
\bigskip}

It is clear from the data that there is
some $Q^2=-q^2$ dependence in the structure functions.
In other words, there are violations of Bjorken scaling
and actually $F_i=F_i(x,Q^2)$. The free parton model
is not completely correct (no big surprise, of course). Our job is to
try to understand these violations in the framework of
QCD.

Let us recall that all the strong interaction effects in
DIS are contained in the hadronic tensor $H_{\mu\nu}$.
For $-q^2\to\infty$ the
integral in eq. 108 is dominated by the $z^2\to 0$ behaviour, so in order
to find
the $q^2$-dependence for large values of $-q^2$ we need to find the
short distance expansion of the product of currents
$J_\mu(z)J_\nu(0)$. We evaluate this expansion in
three steps$^{31)}$

\noindent
$\bullet$ We decompose
$$\eqalign{J_\mu(z)J_\nu(0) &=(\partial_\mu\partial_\nu-
g_{\mu\nu}\partial^2) O_L(z,0)\cr
&+ (g_{\mu\lambda}\partial_\rho\partial_\nu
+g_{\rho\nu}\partial_\mu\partial_\lambda
-g_{\mu\lambda} g_{\rho\nu}\partial^2
-g_{\mu\nu}\partial_\lambda\partial_\rho)O_2^{\lambda\rho}(z,0)\cr
&+\dots\cr}\eqno(116)$$
$O_2$ is assumed to be symmetric in its indices. The dots stand
for terms which are antisymmetric in $\mu\nu$, which we will
not take into account (they would contribute to the structure
function $F_3$; we will not consider it here).
Current conservation is built in this expression, which is
otherwise completely general.

\noindent
$\bullet$ We expand eq. 116 in
a complete basis of local operators using
the Operator Product Expansion (OPE)$^{32)}$.
$$O_2^{\lambda\rho}(z,0)=\sum c^i_{2,n}(z^2)z^{\mu_1}
\dots z^{\mu_n} O^{i\,\lambda\rho}_{2\,\mu_1\dots\mu_n}(0),\eqno(117)$$
$$O_L(z,0)=\sum c^i_{L,n}(z^2)z^{\mu_1}\dots z^{\mu_n}
O^i_{L\,\mu_1\dots\mu_n}(0).\eqno(118)$$
There may be more than one operator with a given set of
quantum numbers at a given order in the expansions eqs. 117-118
and we have included an index $i$.
The functions $c^i_n$ are called Wilson coefficients.
Both sides of eqs. 117-118 must agree when
inserted in any expectation value or Green function. On dimensional
grounds,
if the dimension of the current $J_\mu$ is $d_0$ and
the one of $O_{\mu_1\dots\mu_n}^i$ is $d_0^i$ then, as
$z^2\to 0$, the Wilson
coefficients $c^i_n$ behave has
$$c^i_n(z^2)\sim (z^2)^{-d_0+{1\over 2}(d_0^i(n)-n)+2}.\eqno(119)$$
The combination $d_0^i(n)-n$ is called the twist
of the operator. For $-q^2\to\infty$ only
$z^2\to 0$ matters, hence we need to retain the
operators of lowest twist, as they provide the
most singular (hence dominant) behaviour. The contribution
of higher twist operators is down by inverse
powers of $Q^2=-q^2$. It is convenient to define the
Fourier transform of the Wilson coefficients
$$ c_n^i(Q^2){2^{n+1}\over {(Q^2)^{n+1}}}q^{\mu_1}\dots
q^{\mu_k}
=\int d^4z e^{iqz}z^{\mu_1}\dots z^{\mu_k}
 c^i_n(z^2).\eqno(120)$$
We have neglected the tensorial structures that would give
less singular contributions in the $z^2\to 0$ limit.
Naively the $c_n^i(Q^2)$ are dimensionless quantities, pure
numbers.
If fact this is not so. The r.h.s. of eqs. 117-118 is
not well defined because when we insert the local
operators in a Green function we will get additional
divergences. We have to renormalize them and this
changes the Wilson coefficients.
 As usual, via the counterterms, which in perturbation theory are a power
series in $\alpha_s$, a logarithmic dependence in a renormalization scale
$\mu$ appears. On dimensional grounds, the dependence must be through
$Q^2/\mu^2$. Here are the scaling violations we were after.

\noindent
$\bullet$  Finally, we need to know the expectation values
of the local operators $O^i_{\mu_1\dots\mu_n}$ in the
proton state
$$\langle P\vert O_{L\,\mu_1\dots\mu_n}^i(0)\vert P\rangle
= A^i_{L,n} P_{\mu_1}\dots P_{\mu_n},\eqno(121)$$
$$\langle P\vert O_{2\,\mu_1\dots\mu_n}^{i\,\lambda\rho}(0)
\vert P\rangle
= A^i_{2,n} P^\lambda P^\rho P_{\mu_1}\dots P_{\mu_n}.\eqno(122)$$
This is certainly not the most general decomposition of these
expectation values, but other tensorial structures
will eventually prove to be of higher twist.
When we collect all the pieces of the calculation and
put everything together, we finally get
$$F_L(n,Q^2)\equiv
\int_0^1 dx x^{n-2}F_L(x,Q^2)=
\sum_i A^i_{L,n}(\mu^2) c^i_{L,n}(Q^2,\mu^2),\eqno(123)$$
$$F_2(n,Q^2)\equiv
\int_0^1 dx x^{n-2}F_2(x,Q^2)=
\sum_i A^i_{L,n}(\mu^2) c^i_{2,n}(Q^2,\mu^2).\eqno(124)$$
Knowing the Wilson coefficients and the expectation values of
the matrix elements we can, in the large $Q^2$ limit, compute
the moments of the structure functions and hence the structure
functions themselves. The scaling violations come about
exclusively through the logarithmic dependence on
$Q^2/\mu^2$ of the Wilson coefficients.

The
product $A_n(\mu^2) c_n(Q^2,\mu^2)$ is $\mu$-independent
because, as we have just seen, is an observable. As
befits a renormalizable theory $c_n(Q^2,\mu^2)$ satisfies
a renormalization-group type equation (we assume
 that there is only one operator of a given dimension
and quantum numbers, so there is no mixing --- we
suppress the $i$ superindex for simplicity)
$$
(\mu{\partial \over {\partial\mu}} +
\alpha_s\beta{\partial\over {\partial\alpha_s}}
-\gamma_{O_n}) c_n(Q^2,\mu^2)=0.\eqno(125)$$
The quantity $\gamma_{O_n}$ is the
anomalous dimension of the operator. It is found by
determining the combination of renormalization constants
that makes it finite.
Eq. 125 can be integrated out
$$c_n(Q^2,\mu^2)=c_n(\mu^2,\mu^2)
\left( {{\alpha_s(\mu^2)}\over {\alpha_s(Q^2)}}
\right)^{{\gamma_{O_n}}\over {\beta_1}}.\eqno(126)$$
Leading to following scaling behaviour for the moments of the
structure functions
$$F_i(n,Q^2)=F_i(n,\mu^2)
\left( {{\alpha_s(\mu^2)}\over {\alpha_s(Q^2)}}
\right)^{{\gamma_{O_n}}\over {\beta_1}},\eqno(127)$$
which is our final expression. Experiments agree
on the whole very nicely with the scaling violations
predicted by QCD. Taking into account all the
subtle points of Quantum Field Theory that have gone
into the analysis, this provides a beautiful
empirical check of the theoretical framework

\vbox{\bigskip \vskip 6truecm
\medskip\centerline{Fig. 27.- The  $Q^2$ evolution predicted
by QCD confronts experiment. From$^{33)}$.}\bigskip}

\bigskip \bigskip
\line{\bf 14.- Altarelli-Parisi Equations\hfil}
\medskip\noindent
Let us now consider an alternative, and perhaps
more appealing, way of deriving the evolution
equation eq. 127. Let us begin by rewriting
it in a differential form. Introducing
the variable
$$t={1\over 2}\log {{Q^2}\over {\Lambda_{QCD}^2}},\eqno(128)$$
we have, for instance,
$${{\partial F_2(n)}\over {\partial t}}
= -{{\gamma_{O_n} \alpha_s(t)}\over {4\pi}}F_2(n).\eqno(129)$$
Let us now introduce the Mellin transform  of the anomalous
dimensions $\gamma_{O_n}$
$$-{1\over 4}\gamma_{O_n}=\int_0^1 dz z^{n-1} P(z).\eqno(130)$$
Using the decomposition
$$F_2(x,t)=x\sum_f \delta_f q_f(x,t)\eqno(131)$$
we arrive at
$$
 {{\partial q_f(x,t)}\over {\partial t}} =
{{\alpha_s(t)}\over \pi}\int_x^1 {{dy}\over y} q_f(y,t)
P({x\over y})
\equiv {{\alpha_s(t)}\over \pi} q_f \otimes P.
\eqno(132)$$
These are the Altarelli-Parisi equations$^{34)}$. They
summarize the rate of change of the parton distribution
functions with $t$.

Let us now try to get a physical picture.
We start by looking at the QCD diagrams that can contribute
to this process. At leading order in $\alpha_s$ we have to
consider the contribution from diagrams
(a)-(e) in fig. 31.
Since we already know that scaling violations are logarithmic
we have to look for possible sources of logs. Calculations
are simplest in an axial gauge. A careful evaluation shows that
ultraviolet divergences
 are absent in this case, so they cannot provide us
with the logs.
Infrared divergences also cancel when all diagrams
are taken into account (although each one of them
is infrared divergent). There is only one
mass singularity, originating from the real gluon emission
diagram (d) that
survives. So this is our only source of logs.

We have then to allow that the parton actually carries
a different momentum fraction $y$ and radiates a
collinear gluon. Since
$F_2$, up to
some kinematical factors, is just a
cross section describable in terms of partons we can
write
$${1\over x} F_2(x,t)=\sum_f \delta_f\int_x^1 dy q_f(y)w_2(x,y,t).
\eqno(133)$$
In the free case (no collinear gluon emitted)
we simply have
$$w_2^{free}={1\over y}\delta({y\over x}-1).\eqno(134)$$

\vbox{\bigskip\vskip 5truecm\medskip
\centerline{Fig. 31.- QCD diagrams contributing to DIS.}
\bigskip}

\noindent
The mass singularity introduces some
 $t$ dependence in $w_2$. Using the
expression of $F_2$ in terms of $q(x)$,
$${{\partial q_f(x,t)}\over {\partial t}}=
\int_x^1 q_f(y){{\partial w_2(x,y,t)}\over {\partial t}}
\eqno(135)$$
We can iterate the process summing up
an infinite ladder of diagrams (fig. 32). This is
accomplished by replacing $q_f(y)$ by
$q_f(y,t)$; the virginal quark is replaced
by a quark that has already exchanged many collinear
gluons.

\vbox{\bigskip\vskip 3truecm\medskip
\centerline{Fig. 32.- `Handbag' and ladder diagrams.}
\bigskip}

We have been considering $F_2$, but the same
procedure can be repeated for any structure function.
As a simplifying hypothesis we have neglected mixing
with other operators. In fact, the evolution equation
is a $(2N_f+1)\times (2N_f+1)$ matrix.
For flavour singlet operators life is
more complicated; there is mixing
with gluon operators and one must also
consider gluon parton distribution functions as well
$$
{{\partial q(x,t)}\over {\partial t}}={{\alpha_s(t)}\over \pi}
\int_x^1 {{dy}\over y}[q(y,t)P_{q\to q}({x\over y})
+g(y,t) P_{g\to q }({x\over y})]\eqno(136)$$
$${{\partial g(x,t)}\over {\partial t}}={{\alpha_s(t)}\over \pi}
\int_x^1 {{dy}\over y}[g(y,t)P_{g\to g}({x\over y})
+q(y,t) P_{q\to g }({x\over y})]\eqno(137)$$
This is schematically depicted in fig. 33. The detailed
form of the Altarelli-Parisi kernels at leading order can
be found in $^{20)}$.
Next to leading expressions for them are
also available in the literature.

The analysis
of Deep Inelastic
Scattering
based either on the OPE or on the Altarelli-Parisis
equations have been
amongst the most clear tests of
perturbative QCD and the best way of determining $\alpha_s$.
However, at present the value of $\Lambda_{QCD}$
extracted from LEP physics is quite competitive, if not better.

\vbox{\bigskip\vskip 3truecm\medskip
\centerline{Fig. 33.- Altarelli-Parisi kernels.}
\bigskip}

\bigskip
\line{\bf 15.- Parton Distribution Functions\hfil}
\medskip\noindent
Although there are some exceptions (notably that of the
pion form factor) we do not know
in general how to compute the
parton distribution functions, even for $-q^2\to\infty$.
Only their evolution can be reliably computed either
through the Operator Product Expansion of the use
of the Altarelli-Parisi equations and this for
large enough values of $-q^2$.

An interesting issue is the behaviour
of the parton distribution functions at the
endpoints $x=0$ and $x=1$. The large
$n$ behaviour of the moments (eq. 123-124) probes
the $x\to 1$ region.
Since it is natural to expect that
at the kinematical boundaries the
parton distribution functions vanish,
one can make the following ansatz
for $x\to 1$
$$q(x,Q^2)\sim A(Q^2) (1-x)^{\nu(\alpha_s(Q^2))-1}\eqno(138)$$
Demanding that eq. 138 fulfills the $q^2$ singlet evolution
equation (136) and (137) leads to
$$A(Q^2)=A_0 {{[\alpha_s(Q^2)]^{-d_0}}\over
{\Gamma(1+\nu(\alpha_s(Q^2)))}}\qquad
\nu(\alpha_s)=\nu_0 -{{16}\over {33-2N_f}}\log\alpha_s(Q^2),
\eqno(139)$$
$$d_0={{16}\over {33-2N_f}}({3\over 4}-\gamma_E)\eqno(140)$$
Likewise, for the gluons we have
$$g(x,Q^2)\sim A_0^\prime {{[\alpha_s(Q^2)]^{-d_0}}\over
{\Gamma(2+\nu(\alpha_s(Q^2)))}}
{{(1-x)^{\nu(\alpha_s(Q^2))}}\over {\log(1-x)}}\eqno(141)$$
The constants $A_0, A^\prime_0$ and $\nu_0$ are not
calculable on perturbative QCD and depend on the
specific operator. $d_0$ is universal.

When $x\to 1$ the gluon distribution functions
approach zero more rapidly than the quark ones. For
large values of $x$ the quark contents of nucleons
is the relevant one. For small values of $x$ the
opposite behaviour takes place, the gluon distribution
function eventually becomes dominant. At
LHC the cross-section
will be greatly dominated by low-$x$ physics and the
important process there will be gluon-gluon scattering. At the
Tevatron the quark contents of protons and antiprotons is still
dominant.

To get a semi-quantitative (but not totally satisfactory)
picture of the low-$x$ behaviour of parton distribution functions,
one inverts eqs. 123-124 via an inverse Mellin transform. Then
$$F_2(x,t)={1\over {2\pi i}}\int dn x^{-(n-1)}F_2(n,t)\eqno(142)$$
The evolution of $F_2(n,t)$ is known. We can write
$$ F_2(n,t)=\exp[f(n,t)]F_2(n,t_0)\eqno(143)$$
Then
we
 proceed to evaluate the r.h.s of eq. 142 by the saddle point
method. Of course this requires that $\log(1/x)$ is large and
this is the reason why this procedure gives only the
small $x$ behaviour. The solution is parametrized in
terms of $F_2(n_0,t_0)$, $n_0$ being the solution
of the saddle point equation for $n$.
Working things out one finds that indeed the
gluon parton distribution function is dominant for low
$x$ behaving as
$$g(x)\sim {1\over x}\exp\sqrt{C(Q^2)\log{1\over x}}\eqno(144)$$
where $C(Q^2)$ is calculable. Unfortunately, this answer is
not totally satisfactory beacuse something must stop the
growth in $g(x)$ for low $x$, or else one runs into unitarity problems
sooner or later, and thus eq. 144 it is not credible all the way to
$x=0$. Technically speaking, there must be corrections that destabilize
the saddle point solution. Physically, the uncontrolled growth of the
gluon distribution is an infrared unstability. The density of soft gluons
is too large. Shadowing and non-linear evolution equations are
the buzzwords here$^{35)}$.

Low $x$ means $-q^2$ large, but fixed, and $\nu\to
\infty$. This is the Regge limit and we understand very little of
QCD is this regime. There are several parametrizations$^{36)}$ of
the quark and gluon distribution functions that fit
the experimental results rather well ($\sim 1$\%) in the
range $1>x>10^{-2}$. These parametrizations are obtained in the following
way. One fits the experimental data at some relatively low value of $Q^2$
using, for instance, a form for the parton distribution function inspired
in the quark model, say $q(x)=A x^a(1-x)^b$. Then one evolves $q(x)$
using the Altarelli-Parisi equations and performs a global fit. The region
below $10^{-2}$
had not been explored experimentally until very recently;
a first look at these low-$x$
values has been provided by the commissioning of HERA.

HERA is a machine ideally suited for an in-depth analysis
of structure functions. Fig. 34 shows the physics reach of
HERA. It should be possible to arrive at very low values
of $x$ (up to $x\sim 10^{-4}$).
There are already very interesting results on the
region $10^{-3}<x<10^{-2}$, which actually show that
most of the parametrizations of the structure
functions perform very poorly when it comes to reproducing
the data at low-$x$. Typically they predict an increase as
$x\to 0$ which is lower than what is actually seen
(fig. 35). The behaviour $F_2(x)\sim x^{-\lambda}$, with
$\lambda\sim 1/2$ as $x\to 0$, which is predicted from the
Kuraev-Fadin-Lipatov evolution equation$^{35)}$ seems to stand
the comparison with HERA results best. However, this behaviour is
still incompatible with unitarity and cannot hold all the way
to $x=0$ either. Obviously there are many open questions in this area.

\vbox{\bigskip\vskip 9truecm\medskip \centerline{Fig. 34.-
The physics reach of HERA. From$^{37)}$.}
\bigskip}

\vbox{\bigskip\vskip 4.5truecm\medskip\centerline
{Fig. 35.- Low $x$ structure functions at HERA$^{38)}$.}
\bigskip}

Complementary information on the
parton distribution functions is provided by the use of sum rules. If
one of the operators that appear in the OPE corresponds to a
conserved current such as, for instance, the non-singlet operator
$$ \bar{u}\gamma_\mu u-\bar{d}\gamma_\mu d.\eqno(145)$$
The corresponding anomalous dimension vanishes and its expectation value
corresponds to a conserved charge measured in the nucleon state$^{39)}$.
Then we know
$$\int dx x^{-1} F^{(NS)}(x)= A_1 c_1 \eqno(146)$$
for non-singlet operators and a linear combination of
$$\int dx F^{(S)}(x)\eqno(147)$$
for the singlet operators. In this case the conserved current
is the energy-momentum tensor. As an example let us
consider $ep$ scattering. The non-singlet conserved
current is $\bar{\psi} \gamma_\mu Q \psi$ ($Q$ being the charge
matrix) and the associated conserved charge is
the nucleon electric charge. Then
$$ \int dx x^{-1} F_2^{ep}(x)={1\over 3}(1+{\cal O}(\alpha_s)).
\eqno(148)$$
Other sum rules are
$$\int dx x^{-1}(F_2^{\bar{\nu}p}(x)-F_2^{\nu p}(x))=0,\eqno(149)$$
due to Adler, or the Gross-Llewellyn-Smith sum rule ($I$: isoscalar
target)
$$\int dx x^{-1} F_3^{\nu I}(x)=3(1+{\cal O}(\alpha_s)).\eqno(150)$$
\bigskip
\line{\bf 16.- Large Rapidity Gaps\hfil}
\medskip\noindent
It would not be appropriate to conclude this review
without saying a few words about some very interesting events
that have been seen at HERA$^{40)}$.

We discussed in section 7 the phenomenon of color coherence. In
DIS we expect that hadron production should take place predominantly in the
color string joining the struck parton and the nucleon remains. In practice
this means that most hadron energy deposition should occur
in an angular region relatively close to the proton beam direction.
Described in terms of the pseudorapidity $\eta=-\log \tan \theta/2$
($\theta$ being the polar angle measured following the incoming
proton) one should expect hadron energy deposition to be at relatively
large values of $\eta$ ($\eta_{max}=4.3$).

While this is the case for the bulk of the event sample (fig. 36),
a small but sizeable fraction remains all the way to
$\eta\sim -2$, with little hadronic activity in between. These
`large rapidity gap' events {\it cannot}
be accounted for in the parton model we have discussed.

It has been suggested that the large rapidity gap events can be explained
in terms of the pomeron$^{41)}$, which is believed to dominate
elactic and driffactive production in hadron-hadron interactions.
Although the pomeron itself falls completely outside perturbative
QCD (presumably you have to sum infinite families of diagrams
to reproduce pomeron-like behaviour), the concept of
pomeron structure functions has been put forward and studied. The
idea is that the pomeron could exhibit a partonic structure
that could be probed in hard diffractive
processes. The events observed ar HERA appear to be
of this kind.

\vbox{\bigskip\vskip 4truecm\medskip\centerline{
Fig. 36.- Rapidity distribution of events}
\bigskip}

\line{\bf 17.- Summary and Outlook\hfil}
\medskip\noindent
QCD is now twenty years old. Obviously we are a long way
from the days when it was hotly debated whether a sensible
Field Theory could possibly be asymptotically free. We know
today that QCD makes perfect sense, in fact QCD is now the paradigm
of a Quantum Field Theory and the primary suspect is QED, which
is believed not to be a consistent theory all the way to zero distances.
This is indeed a paradoxical situation.

Quantum Chromodynamics is a strongly coupled theory. Ultimately this
can be traced to the fact that $\Lambda_{QCD}$ is similar
in magnitude to the light quark masses. For very heavy quarks perturbation
theory gives reasonable estimates, but fails for light quarks where
confinement plays a crucial role.

Then it would have seemed rather hopeless twenty years ago  to expect that
we would be able to measure $\alpha_s$ with a mere $5\%$ error. Confinement
makes most hadronic observables to depend on $\alpha_s$ in an unknown
form. Non-perturbative studies such as lattice QCD can, in principle,
unveil the $ \alpha_s$ dependence, but we are still a long way
from reaching the above precision, even though these studies provide
overwhelming evidence
 that QCD does describe the
hadronic world.

Although many valuable theoretical developments exist by now, it is
through experiment that we have ultimately learnt most of what we know about
QCD and $\alpha_s$. At LEP we are in the privileged situation of
having an experiment with high statistics, a clean set-up and typical
processes with a large momentum transfer, leading to a reasonable
convergence of perturbative expansions for inclusive
processes. Thanks to this triple conjunction we have been able
to reach the level of accuracy of a few per cent.

Yet, with practically everything computed at the next-to-leading
order and with no other experiment of similar quality in sight, it
is hard to imagine a sizeable reduction in the error of $\alpha_s$
in the near future.

HERA will probably capture most of the interest of workers in the
field in years to come. There we have the possibility of exploring a new
kinematic range in Deep Inelastic Scattering. Sensitivity to the
low-$x$ range in structure functions will allow us to explore the
region where parton wave functions begin to overlap and
perturbation theory fails to describe the physics even when
$Q^2\to\infty$.

A spectacular evidence for the insufficiency of perturbation
theory even for processes with large $Q^2$ is to be seen in the detection of
large rapidity
gap events, interpreted in terms of collective modes of dual theory. This
is not really new; Regge theory has been long required to understand the
elastic and low $p_T$ behaviour of hadron interactions. What is
new and interesting is that at HERA we will probably be able to
interpolate smoothly between the Regge limit of QCD and
ordinary perturbation theory. Many surprises are probably
in store for us.

\bigskip
\line{\bf Acknowledgements\hfil}
\medskip\noindent
These lectures were written up during the author's visit to the
Theory Group of Fermi National Laboratory whose hospitality is gratefully
acknowledged. We thank J.Fuster and Ll.Garrido for several discussions
concerning the determination of $\alpha_s$. Thanks are also due to
R.Tarrach for reading the manuscript, although the author is solely
responsible of any remaining mistakes. The
financial support of CICYT grant AEN93-0695 and CEE grant CHRX CT93 0343 is
gratefully acknowledged.

\bigskip
\line{\bf Bibliography\hfil}
\medskip

\item{1)}{See for instance: F.J.Yndur\'ain,{\it The Theory of Quark
and Gluon Interactions}, Springer Verlag; G.Sterman, {\it Quantum
Field Theory}, Cambridge University Press, T. Muta, {\it Foundations
of Quantum Chromodynamics}, World Scientific.}
\item{2)}{J.Fuster, these Proceedings.}
\item{3)}{D.Adams et al. (the SMC collaboration), CERN-PPE-94-59;
For theoretical discussions see for instance: G. Altarelli,
in Proceedings of the International School of Subnuclear Physics, Erice
1989; A. Manohar, in Proceedings of the Polarized Collider Workshop,
University Park, 1990.}
\item{4)}{See for instance: E.Laenen et al, Phys. Rev.D49 (1994) 5753,
and references therein.}
\item{5)}{M.Aguilar-Benitez et al. (the Particle Data Group), in
Review of Particle Properties, Phys. Rev. D45 (1992) S1.}
\item{6)}{In addition to the references listed in $^{1)}$ see
also: J.D.Bjorken and S.Drell, {\it Relativistic Quantum Fields},
Mc.Graw-Hill; C.Itzykson and J.B. Zuber, {\it Quantum Field Theory},
Mc-Graw-Hill.}
\item{7)}{See for instance: R. Jackiw in {\it Current Algebra
and Anomalies}, Princeton.}
\item{8)}{For a simple exposition see: S.Coleman, in {\it Aspects of
Symmetry}, Cambridge University Press.}
\item{9)}{For a more detailed discussion and further references see: E. de
Rafael, in Proceedings of the GIFT Workshop on Quantum Chromodynamics, Jaca,
1979, Springer Verlag.  }
\item{10)}{M.Gell-Mann and Y.Ne'eman, in {\it The Eightfold Way}, Benjamin.}
\item{11)}{See e.g.: S.Coleman, in $^{8)}$}
\item{12)}{T.Hebbeker, Physics Reports 217 (1992) 69}
\item{13)}{Although in the context of weak interactions a clear discussion
is given in: J.C.Taylor, in {\it Gauge Theories of Weak Interactions},
Cambridge University Press}
\item{14)}{C.Callan, Phys. Rev. D2 (1970) 1541; K.Symanzik, Com. Math.
Phys. 18 (1970) 227; see also $^{9)}$ }
\item{15)}{For a discussion and further references: F.J.Yndur\'ain in
$^{1)}$; see also S.Weinberg, Phys. Rev. D8 (1973) 3497.}
 \item{16)}{F.Wilczek, in Proceedings of the Lepton-Photon
Conference, Ithaca, 1993.}
\item{17)}{For a recent comprehensive work see: S.Titard and
F.J.Yndur\'ain, Phys. Rev D49 (1994) 6007 and FTUAM-94-6.}
\item{18)}{H. Georgi, in {\it Weak Interactions and Modern Particle
Theory}, Benjamin-Cummings.}
\item{19)}{H.Leutwyler, Phys. Lett. 98B (1981) 447.}
\item{20)}{See e.g.: R.K.Ellis and W.J.Stirling,
Lectures given at the CERN School of Physics, FERMILAB-Conf-90/164-T.}
\item{21)}{J.Donoghue, E.Golowich and B. Holstein, in {\it
Dynamics of The Standard Model}, Cambridge University Press.}
\item{22)}{See e.g.: A.C. Mattingly and P.M.Stevenson, Phys. Rev.
D49 (1994) 437.}
\item{23)}{E. Braaten, S.Narison and A.Pich, Nucl. Phys. B373 (1992) 581.}
\item{24)}{F.Bloch and A. Nordsieck, Phys. Rev. 52 (1937) 54.}
\item{25)}{T.Kinoshita, J. Math. Phys. 3 (1962) 650;
T.D.Lee and M.Nauenberg, Phys. Rev 133B (1964) 1549.}
\item{26)}{T.Appelquist et al. Phys. Rev. Lett. 36 (1976) 768;
Nucl. Phys. B120 (1977) 77. See also G.Sterman in $^{1)}$.}
\item{27)}{N.Magnoli, P.Nason and R.Rattazzi, Phys. Lett. B252 (1990) 271.}
\item{28)}{G.Miller et al. (the SLAC-MIT collaboration), Phys. Rev.
D5 (1972) 528.}
\item{29)}{See e.g. ref $^{20)}$ or F.J.Yndur\'ain in
$^{1)}$.}
\item{30)}{M.Derrick et al. (the ZEUS collaboration), Phys. Lett.
B315 (1993) 481.}
\item{31)}{Our discussion follows T. Muta in ref $^{1)}$.}
\item{32)}{K.Wilson, Phys. Rev. 179 (1969) 1499. For a detailed practical
example see: P.Pascual and R.Tarrach in {QCD: Renormalization for the
Practitioner}, Springer Verlag. }
\item{33)}{A.C.Benvenuti et al., Phys. Lett. B223 (1989)485.}
\item{34)}{G.Altarelli and G.Parisi, Nucl. Phys. B126 (1977) 298.}
\item{35)}{E.Kuraev, L.Lipatov and V.Fadin, Sov. Phys. JETP 45 (1977) 199;
see also E.Levin, in Proceedings of the Blois Conference on Elastic and
Diffractive Scattering, Providence, 1993}
\item{36)}{A.Martin, W.J.Stirling and R.Roberts, Phys. Rev. D 47 (1993) 867}
\item{37)}{G.A.Schuler, in Physics at HERA, Vol. 1}
\item{38)}{T. Ahmed et al (the H1 collaboration), Nucl. Phys. B407 (1993)
515; Phys. Lett. B321 (1994) 161}
\item{39)}{See $^{1)}$. See also: A.J.Buras, Rev. Mod. Phys. 52 (1980) 199}
\item{40)}{L.Labarga, these Proceedings; M.Derrick et al. (the ZEUS
collaboration), Phys. Lett. B332 (1994) 228; T.Ahmed et al. (the H1
collaboration), DESY-94-133}
\item{41)}{See e.g. A.Donnachie and P.V.Landshoff, Nucl. Phys. 231 (1984)
189; Nucl. Phys. B267 (1986) 690 }
\item{42)}{G.Ingelman and P.Schlein, Phys. Lett. B152 (1985) 256; E. Berger
et al., Nucl. Phys. B 286 (1987) 704}
\bye